\documentclass[journal]{IEEEtran}
\usepackage{cite}
\usepackage[pdftex]{graphicx}
\usepackage{amsmath}
\usepackage{amssymb}
\usepackage{mathtools}
\usepackage{color}
\usepackage{threeparttable}

\newcommand{\wk}{\text{\tiny WK}}

\newcommand{\rev}[1]{{#1}}

\begin{document}

\title{Noise Filtering and Prediction in Biological Signaling
  Networks}

\author{
    \IEEEauthorblockN{David Hathcock, James Sheehy, Casey Weisenberger, Efe Ilker, and Michael Hinczewski}\\
    \IEEEauthorblockA{Department of Physics, Case Western Reserve University, OH 44106}
}

\maketitle

\begin{abstract}
  Information transmission in biological signaling circuits has often
  been described using the metaphor of a noise filter.  Cellular
  systems need accurate, real-time data about their environmental
  conditions, but the biochemical reaction networks that propagate,
  amplify, and process signals work with noisy representations of that
  data.  Biology must implement strategies that not only filter the
  noise, but also predict the current state of the environment based
  on information delayed due to the finite speed of chemical
  signaling.  The idea of a biochemical noise filter is actually more
  than just a metaphor: we describe recent work that has made an
  explicit mathematical connection between signaling fidelity in
  cellular circuits and the classic theories of optimal noise
  filtering and prediction that began with Wiener, Kolmogorov,
  Shannon, and Bode.  This theoretical framework provides a versatile
  tool, allowing us to derive analytical bounds on the maximum mutual
  information between the environmental signal and the real-time
  estimate constructed by the system.  It helps us understand how the
  structure of a biological network, and the response times of its
  components, influences the accuracy of that estimate.  \rev{The
    theory also provides insights into how evolution may have
    tuned enzyme kinetic parameters and populations to optimize
    information transfer.}
\end{abstract}

\section{Introduction}

In the acknowledgments of his seminal 1948 paper, {\it A Mathematical
  Theory of Communication}~\cite{Shannon48}, Claude Shannon left an
interesting clue to the evolution of his ideas on information:
``Credit should also be given to Professor N. Wiener, whose elegant
solution of the problems of filtering and prediction of stationary
ensembles has considerably influenced the writer’s thinking in this
field.''  Seven years earlier, both Shannon, a newly minted
Ph.D. beginning his career at Bell Labs, and Norbert Wiener, the
already legendary mathematical prodigy of MIT, were assigned to the
same classified American military program: designing anti-aircraft
fire-control directors, devices that could record and analyze the path
of a plane carrying out evasive maneuvers, and then aim the gun to a
predicted future position.  Though the project had a very specific
technical aim, Wiener conceived of it more broadly, as part of a
general class of problems where a signal (in this case the time series
of recorded plane positions) contained an underlying message (the
actual positions) corrupted by noise (inevitable tracking errors).  To
filter out the noise and optimally predict a future location, one
needed a new mathematical framework (and in particular a statistical
theory) for communications and control systems.  During the duration
of the project, Shannon traveled every few weeks to MIT for meetings
with Wiener~\cite{Kline15}, and their discussions were the germ of
distinct intellectual trajectories that converged in 1948 with the
publication of two founding texts of this new framework: Shannon's
paper in July, and Wiener's magnum opus, {\it
  Cybernetics}~\cite{Wiener48}, in October.  Both works introduced a
probabilistic measure for the amount of information in a signal: the
entropy (or ``negentropy'' in Wiener's case, differing from Shannon's
definition by a minus sign), borrowed from statistical physics.  Thus
from the very beginnings of information theory as a discipine,
quantifying information and optimizing its transmission have been
intimately linked.

Wiener's classified wartime report on optimal noise filtering and
prediction, the notoriously complex mathematical tour de force that
stimulated Shannon, was made public in 1949~\cite{Wiener49}.  The
following year, Shannon and Henrik Bode provided a remarkably simple
reformulation of Wiener's results~\cite{Bode50}, a paper that is
itself a small masterpiece of scientific exposition.  This has since
become the standard description of Wiener's theory and the start of
an explosion of interest in applying and generalizing noise filter
ideas~\cite{Kailath74}.  As with other areas of information and
control theory, the breadth of applications is striking.  The noise
filter story that began with anti-aircraft directors eventually led,
through Kalman and Bucy~\cite{Kalman60,Kalman61}, to the navigation
computers of the Apollo space program.  As a practical achievement,
guiding spacecraft to the moon was a dramatic turnaround from the
inauspicious beginnings of the field. The anti-aircraft device
designed by Wiener and Julian Bigelow, based on Wiener's theory, was
considered too complicated to produce during wartime and the project
was terminated in 1943, with research redirected toward simpler
designs~\cite{Kline15}.  Mathematical triumphs do not always translate
to marvels of engineering.

In recent years, noise filter theory has found a new arena far from its
original context: as a mathematical tool to understand the constraints
on transmitting information in biological signaling
networks~\cite{Hinczewski14,Becker15,Hinczewski16,Zechner16}.  The
imperative to deal with noise is particularly crucial for living
systems~\cite{Altschuler10}, where signals are often compromised by
stochastic fluctuations in the number of molecules that mediate the
transmission, and the varying local environments in which signaling
reactions occur~\cite{Becskei00,Thattai01,Paulsson04,Cheong11}.  While
noise can be beneficial in special cases~\cite{Cai08}, it is typically
regulated by the cell to maintain function, as evidenced by suppressed
noise levels in certain key proteins~\cite{Newman06}.  Several natural
questions arise: what mechanisms does the cell have at its disposal to
decrease noise?  Given the biochemical expense of signaling, can cells
deploy optimal strategies that achieve the mathematical limits of
signaling fidelity?  Do we know explicitly what these limits are for a
given biochemical reaction network?

In this paper, we will review recent progress in tackling these
questions using noise filter ideas.  Our main focus will be Wiener's
solution for the optimal filter.  This is now often called the
Wiener-Kolmogorov (WK) filter since Andrei Kolmogorov in the Soviet
Union worked out the discrete, time-domain version of the
solution~\cite{Kolmogorov41} independently in 1941, just as Wiener was
completing his classified report.  We will use the Bode-Shannon
formulation of the problem~\cite{Bode50}, which provides a convenient
analytical approach.  We illustrate the versatility of the WK theory
by applying it to three examples of biochemical reaction networks.
Two interesting facets of the theory emerge: i) Biology can implement
optimal WK noise filters in different ways, since what constitutes the
``signal'', the ``noise'', and the ``filter'' depend on the structure
and function of the underlying network. ii) Because of the finite
propagation time of biological signals, often relayed through multiple
stages of chemical intermediates, cellular noise filters are
necessarily predictive.  Any estimate of current conditions will be
based on at least somewhat outdated information.  Before we turn to
specific biochemical realizations of the WK filter, we start by giving
a brief overview of the basic problem and the generic form of the WK
solution.

\section{Overview of WK filter theory}

\subsection{Noise filter optimization problem}\label{nfop}

Imagine we have a stochastic dynamical system which outputs a
corrupted signal time series $c(t) = s(t) + n(t)$, where $s(t)$ is the
underlying ``true'' signal and $n(t)$ is the noise.  We would like to
construct an estimate $\tilde{s}(t)$ that is as close as possible to
the signal $s(t)$, in the sense of minimizing the relative mean
squared error:
\begin{equation}\label{e1}
\epsilon(s(t),\tilde{s}(t)) = \frac{\langle (\tilde{s}(t) - s(t))^2 \rangle}{\langle s^2(t) \rangle},
\end{equation}
where the brackets $\langle\, \rangle$ denote an average over an
ensemble of stochastic trajectories for the system.  We assume the
system is in a stationary state, which implies that $\epsilon$ is
time-independent.  The time series are defined such that the mean
values
$\langle s(t) \rangle = \langle \tilde{s}(t)\rangle = \langle
n(t)\rangle = 0$. \rev{A fundamental property of linear systems is
  that they can be expressed using a convolution, and in this case the
  estimate $\tilde{s}(t)$ is the convolution of a filter function
  $H(t)$ with the corrupted signal}:
\begin{equation}\label{e2}
\tilde{s}(t) = \int_{-\infty}^\infty dt^\prime\,H(t-t^\prime) c(t^\prime).
\end{equation}
\rev{Eq.~\eqref{e2} constitutes what we will refer to as a linear noise
filter.}  For the biological systems we discuss below, $s(t)$,
$\tilde{s}(t)$, and $n(t)$ will depend on the trajectories of
molecular populations, which in turn will depend on the system
parameters.  If $s(t)$ and $\tilde{s}(t)$ can be related as in
Eq.~\eqref{e2}, and minimizing the difference between $\tilde{s}(t)$
and $s(t)$ is biologically important, then we say that our system can
be mapped onto a linear noise filter described by $H(t)$.  Changing
the system parameters will change $H(t)$, and the big question we
would like to ultimately answer is whether a particular system can
approach optimality in noise filtering.

Filter optimization means finding the function $H(t)$ that minimizes
$\epsilon$.  What makes this problem non-trivial is that $H(t)$ cannot
be completely arbitrary.  Physically allowable filter functions obey a
constraint of the form: $H(t) = 0$ for all $t < \alpha$, where
$\alpha \ge 0$ is a constant.  For $\alpha = 0$, this corresponds to
enforcing causality: if $\tilde{s}(t)$ is being constructed in real
time (as is the case for the biological systems we consider below) it
can only depend on the past history of $c(t^\prime)$ up to the present
moment $t^\prime = t$.  If $\alpha >0$, the estimate is constrained
not only by causality, but by a finite time delay.  Only the past
history of $c(t^\prime)$ up to time $t^\prime = t-\alpha$ enters into
the calculation of $\tilde{s}(t)$.  The two cases of $\alpha$ are
referred to as pure causal filtering ($\alpha = 0$) and causal
filtering with prediction ($\alpha >0$), since the latter attempts to
predict the value of $s(t)$ from data that lags an interval $\alpha$
behind~\cite{Bode50}.

As defined in Eq.~\eqref{e1}, the error $\epsilon$ is in the range
$0 \le \epsilon < \infty$.  For any given filter function $H(t)$, we
can carry out the rescaling $H(t) \to A H(t)$, where $A$ is a
constant.  This transforms $\tilde{s}(t) \to A \tilde{s}(t)$.  The
value of the scaling factor $A$ that minimizes $\epsilon$ is
$A = \langle \tilde{s}(t) s(t) \rangle / \langle {\rev{\tilde{s}}}^2(t) \rangle$, and
we denote the resulting value of $\epsilon$ as $E$:
\begin{equation}\label{e3}
E = \text{min}_A \,\epsilon(s(t),A \tilde{s}(t)) = 1 - \frac{\langle \tilde{s}(t) s(t) \rangle^2}{\langle s^2(t) \rangle \langle \tilde{s}^2(t)\rangle}.
\end{equation}
This alternative definition of error always lies in the range
$0 \le E \le 1$, and is independent of any rescaling of either
$\tilde{s}$ or $s$.  Minimizing $\epsilon$ over all allowable $H(t)$
is equivalent to minimizing $E$, and in fact $E=\epsilon$ for the
optimal $H(t)$.  We will choose $E$ as the main definition of error in
our discussion below.

Before describing the solution for the optimal $H(t)$, it is worth
noting that $E$ (or $\epsilon$) are not the only possible measures of
similarity between $s(t)$ and $\tilde{s}(t)$.  If the the values
$s(t)$ and $\tilde{s}(t)$ at any given time $t$ in the stationary
state have a joint probability distribution
${\cal P}(s(t),\tilde{s}(t))$, then another commonly used measure is
the mutual information~\cite{Levine07,Mugler09,Walczak09}:
\begin{equation}\label{e4}
I\rev{(s;\tilde{s})} = \int ds \int d\tilde{s}\, {\cal P}(s,\tilde{s}) \log_2 \frac{{\cal P}(s,\tilde{s})}{{\cal P}(s) {\cal P}(\tilde{s})},
\end{equation}
where ${\cal P}(s) = \int d\tilde{s}\,{\cal P}(s,\tilde{s})$ and
${\cal P}(\tilde{s}) = \int ds\,{\cal P}(s,\tilde{s})$ are the
marginal stationary probabilities of  $s$ and $\tilde{s}$
respectively.  When $s$ and $\tilde{s}$ are completely uncorrelated,
${\cal P}(s,\tilde{s}) = {\cal P}(s) {\cal P}(\tilde{s})$, resulting
in $I = 0$, the minimum possible value, with the corresponding value
of error being $E = 1$.  Non-zero correlations between $\tilde{s}$ and
$s$ yield $I >0$, and $E <1$.

There is one special case where $I$ and $E$ have a simple analytical
relationship~\cite{Becker15}.  If ${\cal P}(s,\tilde{s})$ is a
bivariate Gaussian distribution of the form
\begin{equation}\label{e5}
{\cal P}(s,\tilde{s}) = \frac{1}{2\pi \sigma \tilde{\sigma} \sqrt{1-\rho^2}} e^{-\frac{1}{2(1-\rho^2)}\left(\frac{s^2}{\sigma^2}+ \frac{\tilde{s}^2}{\tilde{\sigma}^2} -\frac{2\rho s \tilde{s}}{\sigma \tilde{\sigma}}\right)},
\end{equation}
then $E = 1 -\rho^2$ and $I = -(1/2) \log_2 E$.  Here
$\sigma = \langle s^2(t)\rangle^{1/2}$ and
$\tilde{\sigma} = \langle \tilde{s}^2(t) \rangle^{1/2}$ are the
standard deviations of the signal and estimate, and
$\rho = \langle \tilde{s}(t) s(t) \rangle / (\sigma \tilde{\sigma})$
is the correlation between them.  The stochastic dynamics in all the
biological examples below will be governed by systems of linear
Langevin equations, and as a result the bivariate Gaussian assumption
of Eq.~\eqref{e5} holds~\cite{Kampen}.  It applies so long as the
Langevin description is valid, namely for large populations that can
be treated as continuous variables.  Thus all our results for minimum
values of $E$ can be directly translated into maximum values of $I$.

\subsection{Optimal WK filter solution}

The optimization problem described above is more convenient to handle
in Fourier space.  The convolution in Eq.~\eqref{e2} becomes:
\begin{equation}\label{e6}
\tilde{s}(\omega) = H(\omega) c(\omega) = H(\omega) ( s(\omega) + n(\omega)),
\end{equation}
where the Fourier transform of a time series $a(t)$ is given by
$a(\omega) = {\cal F}[a(t)] \equiv \int_{-\infty}^{\infty} dt\, a(t)
\exp(i\omega t)$.  The error $E$ in Eq.~\eqref{e3} can be expressed as:
\begin{equation}\label{e7}
E = 1 - \left.\frac{\left({\cal F}^{-1}[H(\omega) P_{cs}(\omega)]\right)^2}{{\cal F}^{-1}[P_{ss}(\omega)] {\cal F}^{-1}[|H(\omega)|^2 P_{cc}(\omega)]}\right|_{t=0},
\end{equation}
where ${\cal F}^{-1}$ is the inverse Fourier transform and
$P_{ab}(\omega)$ is the cross-spectral density for series $a(t)$ and
$b(t)$, defined through the Fourier transform of the correlation
function
\begin{equation}\label{e8}
P_{ab}(\omega) = {\cal F}[\langle a(t+t^\prime) b(t^\prime)\rangle],
\end{equation}
or equivalently through the relation
$\langle a(\omega) b (\omega^\prime)\rangle = 2\pi P_{ab}(\omega)
\delta(\omega + \omega^\prime)$.  The constraint $H(t) = 0$ for
$t< a$ in all physically allowable filter functions translates in
Fourier space into the fact that the function
${\cal F}[H(t+\alpha)] = e^{-i \omega \alpha} H(\omega)$ must be ``causal'' in the following
sense~\cite{ChaikinLubensky}: when extended to the complex $\omega$
plane, it can have no poles or zeros in the upper half-plane
$\text{Im}\, \omega >0$.

Finding the $H(\omega)$ that minimizes $E$ among this restricted class
of filter functions yields the WK optimal filter
$H_\wk(\omega)$~\cite{Wiener49,Kolmogorov41}.  The end result,
expressed in the form worked out by Bode and Shannon~\cite{Bode50},
is:
\begin{equation}\label{e9}
H_\wk(\omega) = \frac{e^{i\omega \alpha}}{P^+_{cc}(\omega)}\left\{\frac{P_{cs}(\omega)e^{-i\omega\alpha}}{P^+_{cc}(-\omega)} \right\}_+,
\end{equation}
where the $+$ superscripts and subscripts refer to two types of causal
decomposition.  The function $P^+_{cc}(\omega)$ is defined by writing
$P_{cc}(\omega) = |P^+_{cc}(\omega)|^2$, where $P^+_{cc}(\omega)$ is
chosen such that it has no zeros or poles in the upper half-plane, and
satisfies $(P^+_{cc}(\omega))^\ast = P^+_{cc}(-\omega)$.  This
decomposition always exists if $P_{cc}(\omega)$ is a physically
allowable power spectrum.  The other decomposition, denoted by
$\{ R(\omega) \}_+$ for a function $R(\omega)$, is defined as
$\{ R(\omega) \}_+ \equiv {\cal F}[\Theta(t) {\cal
  F}^{-1}[R(\omega)]]$, where $\Theta(t)$ is a unit step
function~\cite{Becker15}.  Alternatively, it can be calculated by
doing a partial fraction expansion of $R(\omega)$ and keeping only
those terms with no poles in the upper half-plane.  If we plug in
$H_\wk(\omega)$ into Eq.~\eqref{e7} for $E$ and carry out the inverse
transforms, we get the minimum possible error for a physically
allowable linear filter, which we denote as $E_\wk$ in the examples
below.

One aspect of the optimal solution should be kept in mind in any
specific application of the filter formalism: $H_\wk$ in
Eq.~\eqref{e9} depends on $P_{cs}$ and $P_{cc}$, and hence $E_\wk$ is
determined once $P_{cs}$, $P_{cc}$, and $P_{ss}$ are given.  For a
particular system, these three spectral densities will depend on some
subset of the system parameters.  Once the densities are fixed,
typically there are remaining parameter degrees of freedom that allow
the filter function $H$ to vary.  However, since these remaining
parameters form a finite set, it is not guaranteed that all possible
functional forms of $H$ are accessible through them.  To make the
system optimal, one should choose these remaining parameters such that
$H = H_\wk$.  In certain cases this can be done exactly, and in other
cases only approximately or not at all.  $E_\wk$ is a lower
bound on $E$ for linear noise filters, but whether it can actually be
reached is a system-specific question.

\section{Realizing noise filters in biological signaling pathways}

\begin{figure}
\centering\includegraphics[width=\columnwidth]{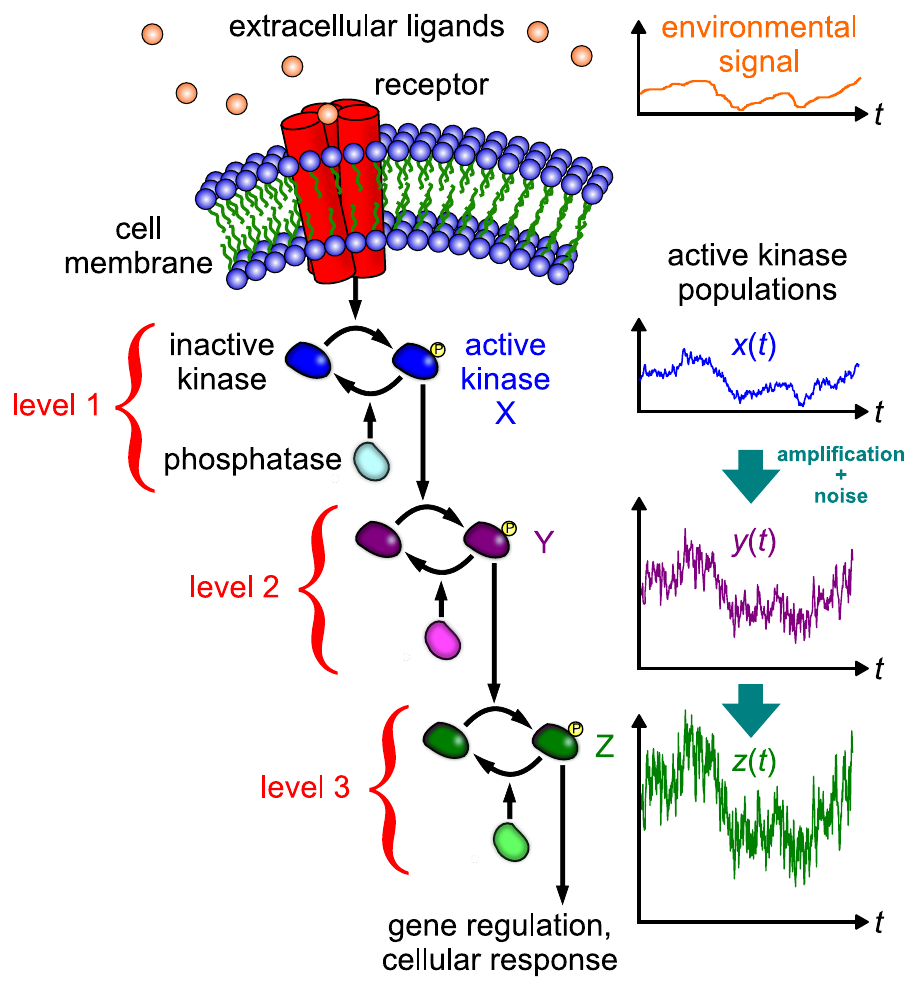}
\caption{\rev{Schematic diagram of a cellular signaling pathway, like
    the MAPK cascade in eukaryotes.  An environmental signal (a
    time-varying concentration of extracellular ligands) is propagated
    through membrane receptors into populations of activated kinase
    proteins.  Each active kinase is turned on through phosphorylation
    reactions catalyzed by a receptor or kinase protein in the level
    above, and turned off through dephosphorylation catalyzed by a
    phosphatase protein.  Since an active kinase can phosphorylate
    many downstream substrates before it is deactivated, the signal is
    amplified as it passes from level to level.  However, because the
    enzymatic reactions are inherently stochastic, noise is introduced
    along with the amplification.}}\label{cas}
\end{figure}

\rev{
  \subsection{Optimizing information transfer in a cellular signaling
    cascade}\label{bio}

  To make use of the definitions of error and mutual information in
  Sec.~\ref{nfop}, we need to translate them into a specific
  biological context.  The first context we will consider
  is a cellular signaling pathway, drawn schematically in
  Fig.~\ref{cas}.  The signal originates in time-varying
  concentrations of external ligand molecules, representing
  environmental factors that are relevant to the cell's functioning.
  These factors include stressors like toxic chemicals or high osmotic
  pressure, or the presence of signaling molecules released by other
  cells (hormones, cytokines, pheromones) that may influence cell
  division, differentiation, and
  death~\cite{Martin05,Mettetal08,Hersen08}.  The signal propagates
  into the cell interior by activation of membrane receptor proteins
  in response to ligand binding.  In order to ensure a robust
  response, which requires a sufficiently large population of active
  proteins at the end of the pathway in order to regulate gene
  expression, the signal is typically amplified through a series of
  enzymatic reactions.  A canonical example of this in eukaryotes is
  the three-level mitogen-activated protein kinase (MAPK)
  cascade~\cite{Martin05}.  Each level involves a substrate protein
  (the kinase) becoming activated through a chemical modification
  (phosphorylation---the addition of one or more phosphate groups)
  catalyzed by an upstream enzyme (a membrane receptor or other
  kinase).  The activated kinase then catalyzes phosphorylation of the
  substrate at the next level down.  The active kinase state is always
  transient, since other protein enzymes (phosphatases) eventually
  catalyze the removal of the phosphate groups, returning the kinase
  to its inactive form.  Thus every substrate is subject to a
  continuous enzymatic ``push-pull'' loop of activation /
  deactivation~\cite{Stadtman77,Goldbeter81,Detwiler00,Heinrich2002}.
  Two features allow for signal amplification: i) during a single
  active interval, a kinase may phosphorylate many downstream
  substrates; ii) the total substrate populations (inactive and
  active) can increase from level to level, for example in a ratio
  like 1:3:6 seen a type of fibroblast~\cite{Sturm10}.  In addition to
  acting like an amplifier, a multi-stage cascade can also facilitate
  more complex signaling pathway topologies, for example crosstalk by
  multiple pathways sharing common signaling
  intermediates~\cite{Saxena99}, or negative feedback from downstream
  species on upstream components~\cite{Sturm10}.

Let us focus for simplicity on a single stage of the cascade, for
example between the active kinase species X and Y shown in
Fig.~\ref{cas}.  Along with amplification, there is inevitably some
degree of signal degradation due to the stochastic nature of the
chemical reactions involved in the push-pull
loop~\cite{Thattai02,TanaseNicola06}.  We can use the formalism of
Sec.~\ref{nfop} to quantify both the fidelity of the transduced
signal and the degree of amplification.  Let us assume the signal is
a stationary time series and hence the kinase populations (in their
active forms) have time trajectories $x(t)$ and $y(t)$ that fluctuate
around mean values $\bar{x}$ and $\bar{y}$.  If
$\delta x(t) = x(t) - \bar{x}$ and $\delta y(t) = y(t) - \bar{y}$ are
the deviations from the mean, the joint stationary probability
distribution ${\cal P}(\delta x(t),\delta y(t))$ allows us to measure
the quality of information transmission from X to Y in terms of the
mutual information $I(\delta x; \delta y)$ defined in Eq.~\eqref{e4}.
Optimization means tuning system parameters (for example enzymatic
reaction constants or mean total substrate / phosphatase
concentrations) such that $I(\delta x; \delta y)$ is maximized.  As
described in the previous section, the tuning is constrained to a
subset of system parameters.  We fix the properties of the input signal
and the added noise due to the enzymatic loop (in the form of the
associated power spectra $P_{ss}$, $P_{cs}$, and $P_{cc}$), and only
vary the remaining parameters.  Let us partition the total set of
system parameters into two parts: the set $\Psi$ which determines the
input and noise, and the remainder $\Omega$.  We will identify these
sets on a case-by-case basis.  Optimization is then seeking the
maximal mutual information over the parameter space $\Omega$:
\begin{equation}\label{a1}
I_\text{max}(\delta x; \delta y) = \text{max}_\Omega I(\delta x;\delta y).
\end{equation}
This formulation means that we are assuming the input signal (which
ultimately arises from some external environmental fluctuations) is
given, but we also fix the degree of noise corrupting the signal.  In
changing $\Omega$, we are looking for the best way to filter out this
given noise for the given input signal.  The result, $I_\text{max}$,
will depend on the input/noise parameters $\Psi$ and we can then
explore what aspects of $\Psi$ determine $I_\text{max}$: are there
particular features of the input signal (or noise corruption) that
make $I_\text{max}$ higher or lower?

This optimization problem becomes significantly easier if
${\cal P}(\delta x,\delta y)$ has the bivariate Gaussian form of
Eq.~\eqref{e5}, which arises if the underlying dynamical system obeys
linear Langevin equations, as mentioned earlier.  The continuous
population approximation, which is a necessary prerequisite of the
Langevin description, is typically valid in signaling cascades, where
molecular populations are large.  Linearization of the Langevin
equations can be validated by comparison to exact numerical
simulations of the nonlinear system~\cite{Hinczewski14}.  If the
approximation is valid, maximizing $I(\delta x; \delta y)$ becomes
mathematically equivalent to minimizing the scale-independent error
$E$ of Eq.~\eqref{e3}, since $I = -(1/2) \log_2 E$.  To make the
connection with the signal $s(t)$ and estimate $\tilde{s}(t)$
explicit, let us define $s(t) \equiv G \delta x(t)$, and
$\tilde{s}(t) \equiv \delta y(t)$, where
\begin{equation}\label{a2}
G \equiv \frac{\langle \delta y^2(t) \rangle}{\langle \delta y(t) \delta x(t)\rangle}.
\end{equation}
This allows Eq.~\eqref{e3} to be rewritten as:
\begin{equation}\label{a3}
\begin{split}
E &= 1 - \frac{\langle \tilde{s}(t) s(t) \rangle^2}{\langle s^2(t) \rangle \langle \tilde{s}^2(t)\rangle} = 1 - \frac{\langle \delta y(t) \delta x(t) \rangle^2}{\langle \delta x^2(t) \rangle \langle \delta y^2(t)\rangle}\\
&=\text{min}_A \epsilon(\delta x(t), A \delta y(t)) =\text{min}_{\tilde{A}} \epsilon(\tilde{A} \delta x(t), \delta y(t)),
\end{split}
\end{equation}
where $\tilde{A} = A^{-1}$, and the last equality follows from the
definition of $\epsilon$ in Eq.~\eqref{e1}.  Thus $G$ in
Eq.~\eqref{a2} is precisely the value of $\tilde{A}$ that minimizes
$\epsilon(\tilde{A} \delta x(t), \delta y(t))$.  In other words we can
interpret $G$ as the amplification factor (or gain~\cite{Detwiler00})
between the deviations $\delta x(t)$ and $\delta y(t)$.  One would
have to multiply $\delta x(t)$ by a factor $G$ in order for the
amplitude of the scaled fluctuations $G \delta x(t)$ to roughly match the
amplitude of $\delta y(t)$.  The gain $G$ is in general distinct from
the ratio of the means, $\bar{y}/\bar{x}$, which could be used as
another measure of amplification.  Note that $G$ and $E$ are defined
through Eqs.~\eqref{a2}-\eqref{a3} for any $\delta x(t)$ and
$\delta y(t)$, whether or not the mutual information
$I(\delta x; \delta y)$ is optimal.  When we tune the system
parameters $\Omega$ such that $I$ reaches its maximum $I_\text{max}$,
the quantities $G$ and $E$ will have specific values.  In the examples
below, optimality will either exactly or to an excellent approximation
coincide with where the system behaves like a WK filter.  We will
denote the specific values of $G$ and $E$ in these cases $G_\wk$ and
$E_\wk$ respectively.}

\begin{figure}
\centering\includegraphics[width=0.65\columnwidth]{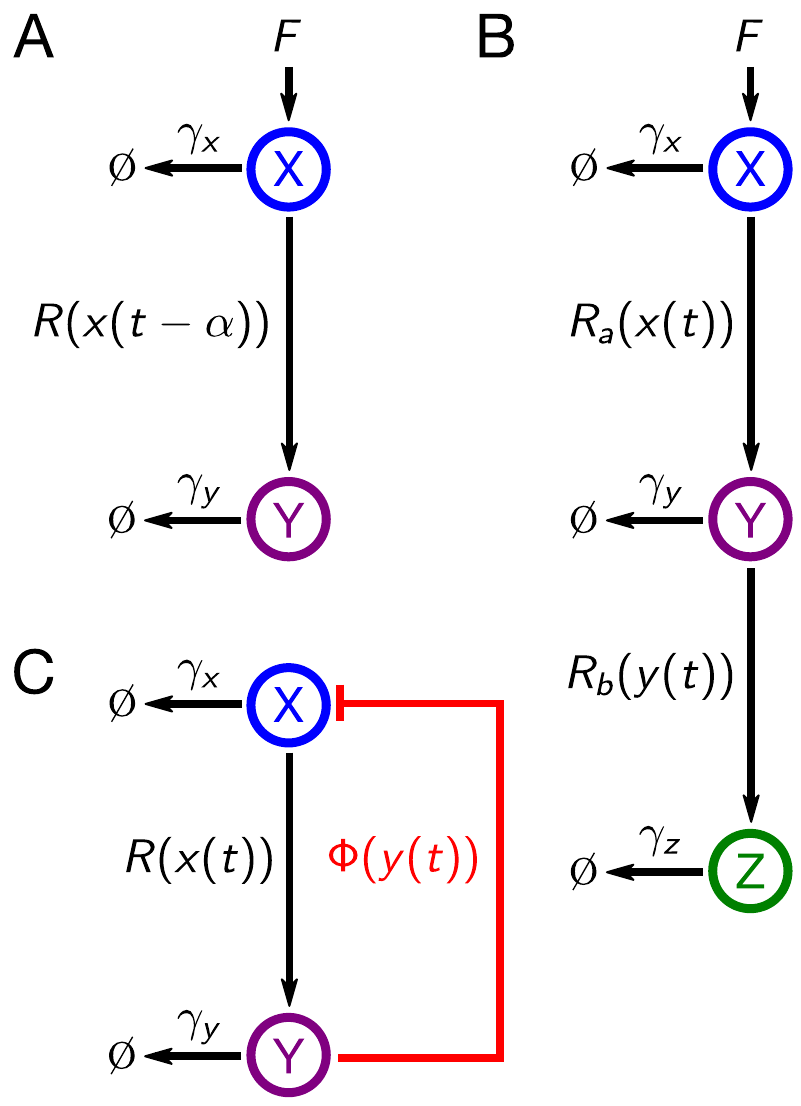}
\caption{Three schematic biochemical reaction networks that
  effectively behave as noise filters.  A: two-species signaling
  pathway with delayed output prodution; B: three-species signaling
  pathway; C: negative feedback loop. The details of the corresponding
  dynamical models are discussed in the text.}\label{net}
\end{figure}

\rev{ We will now show how the filter theory can be applied to two
  simple reaction systems motivated by signaling pathways, illustrated
  in Fig.~\ref{net}A-B.  The simplest case (Fig.~\ref{net}A) is a
  two-species signaling system like the X and Y case described above,
  which could represent a single stage in a signaling pathway.
  Alternatively, one could interpret this system as a coarse-grained
  approximation of a more complex pathway, explicitly considering only
  the first and last species, and with propagation through
  intermediate species modeled as a time-delayed production function.
  The second example (Fig.~\ref{net}B) illustrates in more detail the
  role of signaling intermediates using a three-species pathway, like
  a MAPK cascade.} In each case we start from a model of the system
dynamics in terms of molecular populations, and then construct a
mapping onto the quantities $s(t)$, $\tilde{s}(t)$, $n(t)$, and $H(t)$
from the filter formalism.  This allows us to explore whether the
system can implement optimal linear noise filtering, with $H(t)$
approaching $H_\wk(t)$.  \rev{Once we understand the conditions for
  optimality, we can also explore how the tuning can occur in specific
  biochemical terms (enzyme populations and kinetic parameters), and
  the potential evolutionary pressures that might drive the system
  toward optimality.}

\subsection{Two-species signaling pathway with time delay}

\subsubsection{Dynamical model}
\rev{Our first example is a minimal model for a single stage in a
  signaling pathway (Fig.~\ref{net}A), outlined in the previous
  section.  An input signal from the environment has propagated up
  through molecular species X, with population $x(t)$, and we will
  investigate the next step in the propagation: activation of a second
  species Y, with population $y(t)$, acting as the output.  Both $x(t)$
  and $y(t)$ will represent kinases in their active (phosphorylated)
  form.  A full description of the enzymatic activation process
  involves multiple reactions and the creation of transient
  intermediate species, for example the formation of bound complexes
  of the enzyme with its substrate.  Our minimal model simplifies the
  activation into a single reaction, though as we discuss later the
  resulting theory holds even for a more detailed biochemical model.
  Since we are not modeling in detail the upstream process by which
  $x(t)$ arises, we need to choose a specific form for the
  time-varying population $x(t)$ which represents the input.  One
  approach which leads to mathematically tractable results is to
  assume $x(t)$ is a stationary Gauss-Markov process~\cite{Becker15}:
  a Gaussian-distributed time trajectory with an exponential
  autocorrelation function,
\begin{equation}\label{a4}
\langle \delta x(t) \delta x(t^\prime) \rangle = (F/\gamma_x)
  \exp(-\gamma_x |t^\prime - t|).
\end{equation}
Thus $x(t)$ is characterized by two quantities: the autocorrelation
time $\gamma_x^{-1}$, which sets the time scale over which the signal
shows significant fluctuations, and the factor $F$, which influences
the scale of the variance, $F/\gamma_x$.  One can generalize the
theory to more complex, non-Markovian forms of the input signal where
the autocorrelation is no longer exponential~\cite{Becker15},
including, for example, a deterministic oscillatory
contribution~\cite{Hinczewski14}.  An effective dynamical model which
yields $x(t)$ with the autocorrelation of Eq.~\eqref{a4} is:
\begin{equation}\label{e10}
\frac{dx(t)}{dt} = F - \gamma_x x(t) + n_x(t).
\end{equation}
Thus, $F$ plays the role of an effective production rate due to
upstream events (probability per unit time to activate X), while
$\gamma_x$ acts like an effective single-molecule deactivation rate
(the action of the phosphatase enzymes turning X off).  We assume a
chemical Langevin description~\cite{Gillespie00} which treats $x(t)$
as a continuous variable.} Stochastic fluctuations are encoded in
the Gaussian white noise function $n_x(t)$, with autocorrelation
$\langle n_x(t) n_x(t^\prime) \rangle = 2 \gamma_x \bar{x}
\delta(t-t^\prime)$.  Here $\bar{x} = F/\gamma_x$ is the mean
population of X \rev{(which is also equal to the variance)}.

The dynamical equation for species Y is analogous, but the activation
rate at time $t$ is given by a production function $R(x(t-\alpha))$
that depends on the population $x(t-\alpha)$ of X at a time offset
$\alpha\ge 0$ in the past.  The interval $\alpha$ represents a finite
time delay for the activation to occur.  For an enzyme catalyzing the
addition of a single phosphate group this delay may be negligible,
$\alpha \approx 0$~\cite{Hinczewski14}, but if the activation process
is more complicated, $\alpha$ could be nonzero.  For example, time
delays might occur if multiple phosphorylation events are necessary to
activate Y (as is the case with many proteins), or if the two species
model is a coarse-grained approximation of a pathway involving many
signaling intermediates.  Here we take $\alpha$ to be a given
parameter, but in the next section we see how such a time delay arises
naturally in a three species signaling cascade, and is related to the
reaction timescale of the intermediate species.  Just as with species
X, there will be enzymes responsible for deactivating Y, with a
corresponding net rate $\gamma_y y(t)$.  The resulting dynamical
equation is:
\begin{equation}\label{e11}
\frac{dy(t)}{dt} = R(x(t-\alpha)) -\gamma_y y(t) + n_y(t).
\end{equation}
The Gaussian white noise function $n_y(t)$ has autocorrelation
$\langle n_y(t) n_y(t^\prime) \rangle = 2 \gamma_y \bar{y}
\delta(t-t^\prime)$, where
$\bar{y} = \langle R(x(t-\alpha)) \rangle / \gamma_y$ is the mean
population of Y.  The final aspect of the model we need to specify is
the form of the production function $R$.  We will initially assume a
linear form, $R(x) = \sigma_0 + \sigma_1 (x-\bar{x})$, with
coefficients $\sigma_0, \sigma_1>0$.  (Later we will consider
arbitrary nonlinear $R$.)  The parameter $\sigma_0$ represents the mean
production rate, $\sigma_0 = \langle R(x(t-\alpha)) \rangle$, while
$\sigma_1$ is the slope of the production function.  The mean Y
population is then $\bar{y} = \sigma_0/\gamma_y$.  Note that for the
linear $R(x)$ to be positive at all $x> 0$, as expected for a rate
function, we need $\sigma_1 \le \sigma_0/\bar{x}$.

\subsubsection{Mapping onto a noise filter} Let us rewrite
Eqs.~\eqref{e10}-\eqref{e11} in terms of the variables
$\delta x(t) = x(t) - \bar{x}$ and $\delta y(t) = y(t) - \bar{y}$
representing deviations from the mean, and transform to Fourier space.
The dynamical system then looks like:
\begin{equation}\label{e12}
\begin{split}
-i\omega \delta x(\omega) &= -\gamma_x \delta x(\omega) + n_x(\omega),\\
-i\omega \delta y(\omega) &= \sigma_1 e^{i\omega\alpha} \delta x(\omega) - \gamma_y \delta y(\omega) + n_y(\omega).
\end{split}
\end{equation}
The \rev{corresponding} solutions for $\delta x(\omega)$ and
$\delta y(\omega)$ are:
\begin{equation}\label{e13}
\begin{split}
\delta x(\omega) &=\frac{n_x(\omega)}{\gamma_x - i\omega},\\
\delta y(\omega) &= \frac{G^{-1}\sigma_1 e^{i\omega\alpha}}{\gamma_y - i\omega} \left(G \delta x(\omega)+ \frac{e^{-i\omega\alpha} G n_y(\omega)}{\sigma_1}\right).
\end{split}
\end{equation}
For the $\delta y(\omega)$ solution we have introduced \rev{a constant
  factor $G$ inside the parentheses}, and divided by $G$ outside the
parentheses.  \rev{This allows us to employ the definitions of $s(t)$
  and $\tilde{s}(t)$ in Sec.~\ref{bio}, with the gain $G$ implicitly
  defined through Eq.~\eqref{a2}.  Eq.~\eqref{e13}} has the same structure as Eq.~\eqref{e6},
with the following mapping:
\begin{align}\label{e14}
\tilde{s}(\omega) &= \delta y(\omega), & s(\omega) &= G \delta x(\omega),\nonumber\\
H(\omega) &= \frac{G^{-1}\sigma_1 e^{i\omega\alpha}}{\gamma_y - i\omega}, & n(\omega) &= \frac{e^{-i\omega\alpha} G n_y(\omega)}{\sigma_1}.
\end{align}
Thus we have a natural noise filter interpretation for the system: for
optimum signal fidelity, we want the output deviations $\delta y$ to
be a scaled version of the input deviations $\delta x$ (with
amplification factor $G$), and hence the estimate $\tilde{s}$ to be as
close as possible to the signal $s$.  The function $H$ plays the role
of a linear noise filter, and $n$ is the noise that occurs in the
transmission, related to the stochastic fluctuations $n_y$ intrinsic
to the production of Y.  So far \rev{we have not written an explicit
  expression for the scaling factor $G$, but through Eq.~\eqref{a2} it
  is a function of the system parameters.  At optimality it will have
  a specific value $G_\wk$ derived below,} corresponding to the
  amplification when the system most closely matches $\tilde{s}$ and
  $s$.
  
The time-domain filter function $H(t)$ is
given by:
\begin{equation}\label{e15}
H(t) = G^{-1}\sigma_1 e^{-\gamma_y(t-\alpha)} \Theta(t-\alpha),
\end{equation}
so it satisfies the constraint $H(t) = 0$ for $t < \alpha$.  Because
of the time delay $\alpha$ in the output production, the filter can
only act on the noise-corrupted signal
$c(t^\prime) = s(t^\prime) + n(t^\prime)$ for $t^\prime < t-\alpha$.
The filter action is a form of time
integration~\cite{Berg77,Bialek05}, summing the corrupted signal over
a time interval $\approx \gamma_y^{-1}$ prior to $t-\alpha$, with the
exponential weighting recent values of the signal more than past ones.

\subsubsection{Optimality}
With the above mapping onto a noise filter, we can now determine the
optimal form $H_\wk(\omega)$, and the associated minimal error
$E_\wk$.  Using Eq.~\eqref{e9} we will optimize $H(t)$ over the class
of all linear filters that satisfy $H(t) = 0$ for $t < \alpha$.  The
spectra needed for the optimality calculation are:
\begin{align}\label{e16}
P_{ss}(\omega) &= \frac{2 G^2 F} {\gamma_x^2+ \omega^2},& P_{nn}(\omega) &= \frac{2 G^2 \sigma_0}{\sigma_1^2},\nonumber\\
P_{cs}(\omega) &= P_{ss}(\omega), & P_{cc}(\omega) &= P_{ss}(\omega) + P_{nn}(\omega),
\end{align}
where we have used $\bar{x} = F/\gamma_x$,
$\bar{y} = \sigma_0/\gamma_y$, and the Fourier-space noise correlation
functions,
\begin{equation}\label{e17}
\begin{split}
\langle n_x(\omega) n_x(\omega^\prime)\rangle &= 4\pi \gamma_x \bar{x} \delta(\omega +\omega^\prime),\\
\langle n_y(\omega) n_y(\omega^\prime)\rangle &= 4\pi \gamma_y \bar{y} \delta(\omega +\omega^\prime),\\
\langle n_x(\omega) n_y(\omega^\prime)\rangle &= 0.
\end{split}  
\end{equation}
\rev{The spectra results of Eq.~\eqref{e16} allow us to identify the
  set $\Psi$ of system parameters that determine the input and noise,
  with the remainder constituting $\Omega$, the set over which we
  optimize.  Note that $P_{ss}$, $P_{cc}$, and $P_{cs}$ explicitly
  depend on every system parameter except $\gamma_y$.  The spectra
  also share the common prefactor $G^2$, which depends on all
  parameters, but this will be canceled out in Eq.~\eqref{e9}, so
  $H_\wk$ and $E_\wk$ will be independent of $G$.  This stems from the
  fact that the $G^{-1}$ factor in $H(\omega)$ of Eq.~\eqref{e15} is
  canceled by the $G$ factors in $s(\omega)$ and $n(\omega)$ in the
  convolution of Eq.~\eqref{e6} for $\tilde{s}(\omega)$.  Thus
  $\Psi = \{F$, $\gamma_x$, $\sigma_0$, $\sigma_1\}$, and there is
  only degree of freedom through which the filter $H(t)$ in
  Eq.~\eqref{e15} can approach optimality, $\Omega = \{ \gamma_y\}$.
  We will return later to the biological significance of tuning
  $\gamma_y$, and its relation to phosphatase populations.}

The first decomposition $P_{cc}^+$ in Eq.~\eqref{e9} is given by:
\begin{equation}\label{e18}
P^+_{cc}(\omega) = \frac{G}{\gamma_x}\left(\frac{2F}{\Lambda}\right)^{1/2}\frac{\gamma_x\sqrt{1+\Lambda} - i\omega}{\gamma_x -i\omega},
\end{equation}
where the dimensionless constant
$\Lambda \equiv \bar{x} \sigma_1^2/(\gamma_x \sigma_0) > 0$.  The
second decomposition in Eq.~\eqref{e9} is:
\begin{equation}\label{e19}
\begin{split}
\left\{\frac{P_{cs}(\omega)e^{-i\omega\alpha}}{P^+_{cc}(-\omega)} \right\}_+ &=  \left\{\frac{e^{-i\omega\alpha}G \gamma_x (2F \Lambda)^{1/2}}{(\gamma_x-i\omega)(\gamma_x\sqrt{1+\Lambda}+i\omega)} \right\}_+\\
&= \frac{e^{-\alpha \gamma_x} G (2 F \Lambda)^{1/2}}{(1+\sqrt{1+\Lambda})(\gamma_x - i \omega)}.
\end{split}
\end{equation}
Plugging Eqs.~\eqref{e18}-\eqref{e19} into Eq.~\eqref{e9} gives the
optimal WK filter:
\begin{equation}\label{e20}
H_\wk(\omega) = \frac{e^{\alpha(i\omega-\gamma_x)}\gamma_x(\sqrt{1+\Lambda}-1)}{\gamma_x\sqrt{1+\Lambda}-i\omega},
\end{equation}
with the corresponding time-domain filter function,
\begin{equation}\label{e21}
H_\wk(t) = e^{-\alpha \gamma_x}(\sqrt{1+\Lambda}-1) \gamma_x e^{-\gamma_x\sqrt{1+\Lambda}(t-\alpha)}\Theta(t-\alpha).
\end{equation}
Comparing Eqs.~\eqref{e15} and \eqref{e21}, we see that
$H(t) = H_\wk(t)$ when the following \rev{condition is fulfilled}:
\begin{equation}\label{e22}
\gamma_y = \gamma_x \sqrt{1+\Lambda}.
\end{equation}
\rev{This} equation relates the timescales $\gamma_y^{-1}$ and
$\gamma_x^{-1}$.  For the time integration of the noise filter to work
most efficiently the integrating time interval $\gamma_y^{-1}$ must be
a fraction $1/\sqrt{1+\Lambda}$ of the characteristic fluctuation time
$\gamma_x^{-1}$ of the input signal.  \rev{When Eq.~\eqref{e22} holds the amplification factor $G = G_\wk$, given by:
\begin{equation}\label{e22b}
G_\wk = \frac{\sigma_1 e^{\alpha \gamma_x}}{\gamma_x(\sqrt{1+\Lambda}-1)}.
\end{equation}}
The minimal
error $E_\wk$ associated with $H_\wk$ is:
\begin{equation}\label{e23}
E_\wk = \frac{2e^{-\gamma_x \alpha}}{1+\sqrt{1+\Lambda}} \left(\cosh(\alpha\gamma_x)+\sqrt{1+\Lambda}\sinh(\alpha\gamma_x)\right).
\end{equation}

\begin{figure}
\includegraphics[width=\columnwidth]{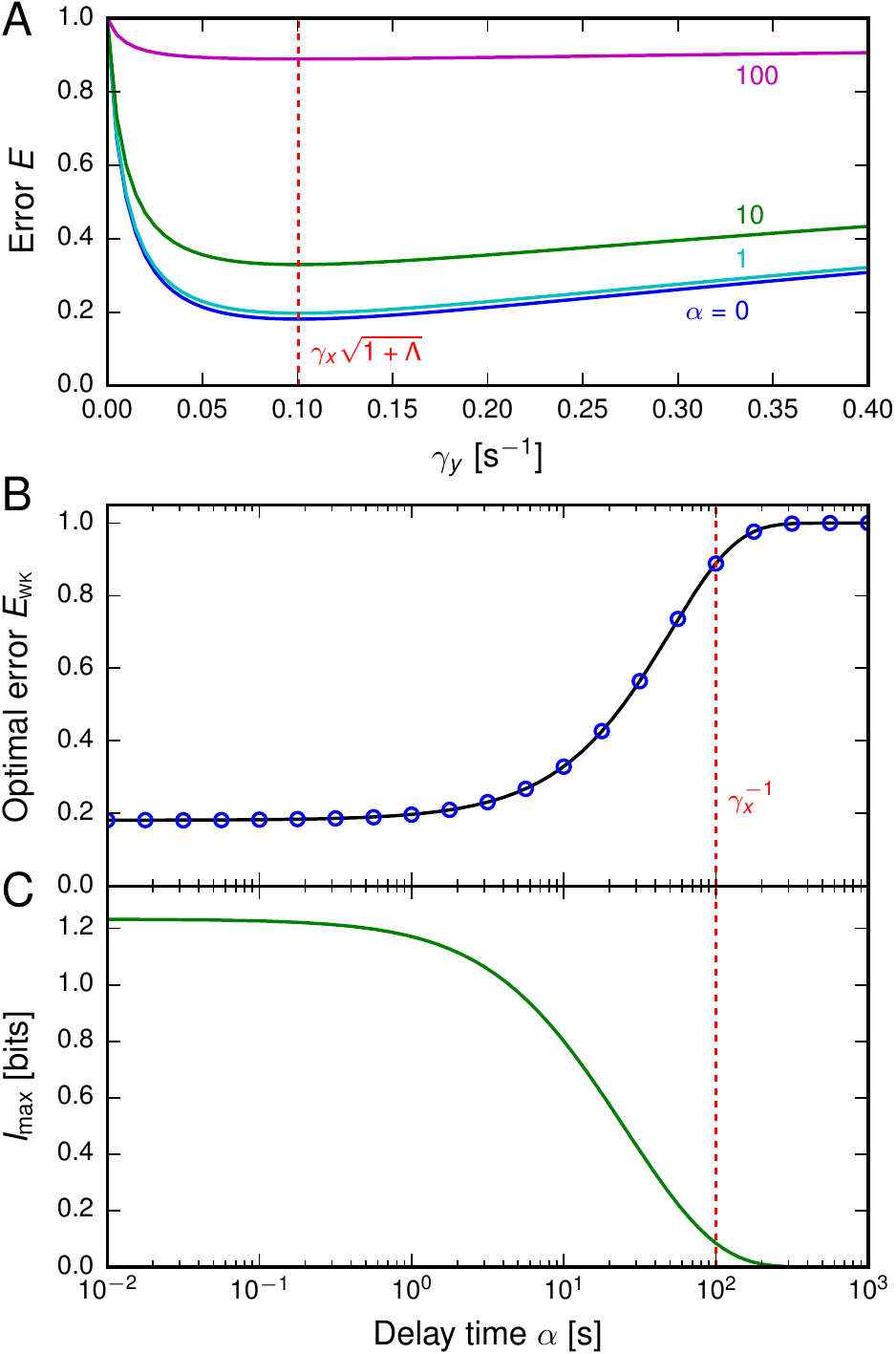}
\caption{Results for the two-species signaling pathway.  A: Error $E$
  versus $\gamma_y$ for time delays $\alpha = 0$, 1, 10, 100 s,
  calculated using Eq.~\eqref{e7}.  The parameters are:
  $\gamma_x = 0.01$ s$^{-1}$, $\sigma_0 = 100$ s$^{-1}$,
  $F = \sigma_1 = 1$ s$^{-1}$.  The dimensionless constant governing
  the filter effectiveness is $\Lambda = 100$.  The dashed vertical
  line shows $\gamma_y = \gamma_x \sqrt{1+\Lambda}$, the location of
  the minimum $E$ where the system behaves like an optimal WK filter.
  B: Optimal error $E_\wk$ versus $\alpha$ for the same parameters as
  panel A. The solid curve is the analytical theory [Eq.~\eqref{e23}]
  while the circles are the results of numerical optimization using
  kinetic Monte Carlo simulations of the system.  The dashed vertical
  line indicates $\gamma_x^{-1}$, the characteristic time scale of
  input signal fluctuations.  C: The corresponding optimal mutual
  information $I_\text{max} = -(1/2) \log_2 E_\wk$ in
  bits.}\label{delay}
\end{figure}

Fig.~\ref{delay}A shows $E$, calculated using Eq.~\eqref{e7}, versus
$\gamma_y$ at several different value of $\alpha$, with the parameters
listed in the caption.  As predicted by the theory, the minimum
$E=E_\wk$ occurs at $\gamma_y = \gamma_x \sqrt{1+\Lambda}$ for all
$\alpha$.  Panel B shows $E_\wk$ versus $\alpha$.  The analytical
curve from Eq.~\eqref{e23} agrees well with numerical optimization
results based on kinetic Monte Carlo simulations~\cite{Gillespie77}.
Panel C shows the corresponding maximum in the mutual information
$I_\text{max} = -(1/2)\log_2 E_\wk$ in bits.  For
$\alpha \gg \gamma_x^{-1}$, the delay is sufficiently large that
prediction based on past data fails, and $E_\wk \to 1$.  In the
opposite regime $\alpha \ll \gamma_x^{-1}$ the optimal error $E_\wk$
from Eq.~\eqref{e23} can be expanded to first order in $\alpha$:
\begin{equation}\label{e25}
E_\wk = \frac{2}{1+\sqrt{1+\Lambda}} + \frac{2\Lambda \alpha\gamma_x}{(1+\sqrt{1+\Lambda})^2} + {\cal O}(\alpha^2).
\end{equation}
The first term on the right is the $\alpha = 0$ optimal error derived
in Ref.~\cite{Hinczewski14}, and the second term is the correction for
small time delays in the signaling pathway.  The first term can be
made arbitrarily small as $\Lambda \to \infty$, with the dimensionless
parameter $\Lambda$ controlling the effectiveness of the noise
filtering.  Since $\Lambda = \bar{x}\sigma_1^2/(\gamma_x \sigma_0)$
and $\sigma_1 \le \sigma_0/\bar{x}$, we have
$\Lambda \le \sigma_0 / F$.  Thus for a given $F$ and $\gamma_x$
(which fixes the input mean $\bar{x} = F/\gamma_x$) we can make
$\Lambda$ large by making the mean production rate $\sigma_0$ of Y as
large as possible, and setting the slope of the production function
$\sigma_1 = \sigma_0/\bar{x}$.  This corresponds to a linear $R$
function of the form $R(x) = \sigma_0 x/\bar{x}$ with no offset at
$x=0$.  Thus better signal fidelity can be bought at the cost of
larger Y production that is also more sensitive to changes in X
(steeper $R$ slope).  Because of the condition
$\gamma_y = \gamma_x \sqrt{1+\Lambda}$, increasing $\Lambda$ also
means a higher rate $\gamma_y$ of Y deactivation to achieve
optimality.

We see that efficient noise filtering is expensive in terms of
biochemical resources.  If activation of Y occurs through the addition
of phosphate groups, each activation event requires hydrolysis of
energy molecules like ATP to provide the phosphates.  And large
$\gamma_y$ requires the cell to maintain sufficiently high populations
of the phosphatase enzymes that remove phosphate groups.  However even
arbitrarily high Y production / deactivation cannot completely
eliminate error when the time delay $\alpha >0$.  We see that the
correction term in Eq.~\eqref{e25} goes to $2\alpha\gamma_x$ as
$\Lambda \to \infty$, and the full expression for $E_\text{wk}$ in
Eq.~\eqref{e23} is bounded from below for all $\Lambda$:
\begin{equation}\label{e26}
E_\wk > 1 - e^{-2\alpha\gamma_x}.
\end{equation}
The optimal prediction filter always incurs a finite amount of error.

\rev{\subsubsection{Phosphatase kinetics and tuning to optimality}

  The theory outlined so far is a minimal model of a signaling system,
  inspired by a kinase-phosphatase push-pull loop.  But does it
  actually capture the behavior of such an enzymatic reaction network
  once we include more details of the chemistry?  And what do the
  conditions for achieving WK optimality tell us about the ways
  evolution may have tuned system parameters to maximize signaling
  fidelity?  To investigate these questions, in
  Ref.~\cite{Hinczewski14} we considered a specific biochemical
  circuit described by the following enzymatic reactions:
\begin{equation}\label{t1}
\begin{split}
K + S
&\xrightleftharpoons[\kappa_u]{\kappa_b} S_K
\xrightarrow{\kappa_r} K + S^\ast\\
P + S^\ast
&\xrightleftharpoons[\rho_u]{\rho_b} S^\ast_P
\xrightarrow{\rho_r} P + S.
\end{split}
\end{equation}
Here $K$ is an active kinase enzyme that can bind to a substrate $S$
and form the complex $S_K$ with reaction rate $\kappa_b$, or unbind
with rate $\kappa_u$.  When the complex is formed a phosphorylation
reaction can occur with rate $\kappa_r$, which for simplicity is
modeled as an irreversible step (since the reverse reaction is highly
unfavorable under physiological conditions).  This reaction releases
the kinase and a phosphorylated substrate, denoted as $S^\ast$.  The
second line in Eq.~\eqref{t1} shows an analogous reaction scheme for
the phosphatase $P$, which can form a complex $S^\ast_P$ with $S^\ast$
and catalyze a dephosphorylation, yielding back the original substrate
$S$.

Because of the binding reactions that form the complexes, this system
is nonlinear and the stochastic dynamics are not analytically
solvable.  However we simulated the dynamics numerically using kinetic
Monte Carlo (KMC)~\cite{Gillespie77} for different choices of the
input signal coming from upstream processes, represented by the free
active kinase population $K$.  To compare directly with the
two-species model, the simplest choice was a Markovian time trajectory
analogous to Eq.~\eqref{a4}, where $K$ has a reaction scheme:
$\emptyset \xrightleftharpoons[\gamma_K]{F} K$.  The effective
activation rate $F$ controls the input fluctuation amplitude, and the
effective deactivation rate $\gamma_K$ controls the fluctuation time
scale $\gamma_K^{-1}$.  In this scenario the numerical simulation
results could be accurately approximated using a mapping onto the
two-species theory outlined above.  The population $x$ in the
two-species model corresponds to the total active kinase population
($K + S_K$), while $y$ corresponds to the the total phosphorylated
substrate ($S^\ast + S^\ast_P$).  The parameters of the two-species
model can be expressed as functions of enzyme kinetic
rates~\cite{Hinczewski14}:
\begin{equation}\label{t2}
\begin{split}
&\gamma_x = \frac{\gamma_K K_M^\text{kin}}{K_M^\text{kin} + [S]}, \quad \gamma_y = \frac{\rho_r [P]}{K_M^\text{pho}+ [P]},\\
&\sigma_1 = \frac{\kappa_r [S]}{K_M^\text{kin}+[S]}, \quad \Lambda = \frac{\kappa_r [S]}{\gamma_K K_M^\text{kin}}.
\end{split}
\end{equation}
For this specific mapping the production function is
$R(x) = \sigma_0 + \sigma_1 (x-\bar{x}) = \sigma_1 x$, since
$\sigma_0 = \sigma_1 \bar{x} = \sigma_1 F/\gamma_x$, and the time
delay $\alpha = 0$ (we do not include activation through multiple
phosphorylations that could delay the signal propagation).  $[S]$ and
$[P]$ are the mean concentrations of substrate and phosphatase in
units of molars (M), and $K_M^\text{kin}$, $K_M^\text{pho}$ are the
Michaelis constants for the kinase and phosphatase respectively, also
in units of M.  These constants are defined as
$K_M^\text{kin} = (\kappa_u + \kappa_r)/k_b$ and
$K_M^\text{pho} = (\rho_u + \rho_r)/\rho_b$ and are commonly used to
characterize enzymatic kinetics~\cite{Fersht}.  The rate of $S^\ast$
production increases with $[S]$, reaches half its maximal value when
$[S] = K_M^\text{kin}$, and approaches the maximum at saturating
substrate concentrations when $[S] \gg K_M^\text{kin}$.  The other
constant $K_M^\text{pho}$ plays the same role for phosphatase: the
rate of $S$ production (from dephosphorylation of $S^\ast$) is
half-maximal when $[S^\ast] = K_M^\text{pho}$.  In deriving
Eq.~\eqref{t2}, we made the following assumptions, justifiable in the
biological context: the enzymes obey Michaelis-Menten kinetics
(catalysis is rate-limiting, so $\kappa_r \ll \kappa_u$,
$\rho_r \ll \rho_u$), and the mean concentration of free active kinase
$[K] \ll [S],[P]$.  Relaxing the Michaelis-Menten assumption leads to
a slightly more complex mapping, but does not significantly alter the
quantitative results below.

In the previous section we showed that for a given input signal and
degree of added noise (i.e. given power spectra $P_{ss}$, $P_{cs}$,
and $P_{ss}$) the two-species system can be tuned to optimality by
varying a single parameter $\gamma_y$.  Maximum mutual information
$I_\text{max}$ is achieved when $\gamma_y$ satisfies the WK condition
of Eq.~\eqref{e22}.  From Eq.~\eqref{t2} we see that $\gamma_y$ is
determined by parameters relating to the phosphatase enzyme: its mean
concentration $[P]$, the catalysis rate $\rho_r$, and its Michaelis
contant $K_M^\text{pho}$.  None of these phosphatase-related
quantities appear in the expressions for the other parameters,
$\gamma_x$, $\sigma_1$, or $\Lambda$.  Thus by altering either the
kinetics or the populations of phosphatase enzymes, biology can tune
the push-pull network to achieve optimal information transfer for a
certain input signal and noise.  $[P]$, $\rho_r$, and $\gamma_y$ are
all possible targets for evolutionary adaptation, but we will focus on
the concentration $[P]$ as the quantity that is most easily modified
(through up- or down-regulation of the genes that express
phosphatases).  Two questions arise, which we will address in turn: i)
is such tuning toward WK optimality plausible in real signaling
pathways given the experimentally-derived data we have on enzyme
kinetics and populations?  ii) What would be the sense of tuning the
system to optimally transmit one type of input signal, characterized
by a certain fluctuation timescale $\gamma_K^{-1}$, since the
environment is likely to provide varying signals with many different
timescales?

\begin{table}

\rev{\begin{threeparttable}[t]
\caption{Substrate/phosphatase concentrations in yeast MAPK signaling pathways, $\gamma_K$ at WK optimality, and $I_\text{max}$}\label{tab1}
\begin{tabular}{|l|cccc|}
\hline
Substrate & $[S]$ ($\mu$M)\tnote{1} & $[P]$ ($\mu$M)\tnote{1,2} & $\gamma_K$ (min$^{-1}$)\tnote{3} & $I_\text{max}$ (bits)\tnote{3}\\
\hline
Hog1 & 0.38 & 1.9 & 3.1 (0.46,\,19) & 1.5 (0.79,\,2.3) \\
Fus3 & 0.47 & 0.081 & 0.60 (0.046,\,6.8) & 2.2 (1.2,\,3.2)\\
Slt2 & 0.18 & 0.081 & 0.43 (0.035,\,4.3) & 1.9 (1.0,\,3.0)\\
\hline
\end{tabular}
\begin{tablenotes}
\item[1] Concentrations are based on protein copy numbers from Ref.~\cite{Ghaemmaghami2003}, using a mean cell volume of 30
fL~\cite{Webster10}.
\item[2] [P] is the total concentration of all phosphatases that target the substrate:  for Hog1 the included phosphatases are Ptc1, Ptc2, Ptc3, Ptp2, Ptp3; for Fus3 they are Ptp2, Ptp3, Msg5; for Slt2 they are Ptp2, Ptp3, Msg5~\cite{Martin05}.
\item[3] Median values, with the 68\% confidence intervals in the
  parentheses, obtained by varying the enzyme kinetic parameters over
  the ranges described in the text.
\end{tablenotes}
\end{threeparttable}}
\end{table}

Let us consider the specific example of yeast MAPK signaling pathways,
and three protein substrates in those pathways that are activated in
response to different environmental signals: Hog1 (osmotic stress),
Fus3 (pheremones), and Slt2 (cell wall damage)~\cite{Martin05}.  Each
substrate is activated by a certain kinase upstream in its MAPK
pathway, and deactivated by a set of phosphatases.  The concentrations
$[S]$ of the substrate and $[P]$ of the phosphatases are listed in
Table~\ref{tab1}.  Since more than one type of phosphatase deactivates
each substrate, $[P]$ is the total concentration of all phosphatases
that share that particular target.  Concentrations are based on
protein copy number measurements from Ref.~\cite{Ghaemmaghami2003},
using a mean cell volume of 30 fL~\cite{Webster10}.  This simple
analysis ignores additional complications like multi-site
phosphyralation of substrates, and the variations in kinetics among
different phosphatase types.  The WK condition in Eq.~\eqref{e22},
when translated into enzymatic parameters using Eq.~\eqref{t2}, can be
solved for $[P]$, implying the following relation at optimality:
\begin{equation}\label{t3}
[P] = \frac{\gamma_K K_M^\text{kin} K_M^\text{pho} \sqrt{1+\kappa_r [S]/(\gamma_K K_M^\text{kin})}}{ \rho_r(K_M^\text{kin}+[S])- \gamma_K K_M^\text{kin} \sqrt{1+\kappa_r [S]/(\gamma_K K_M^\text{kin})}}.
\end{equation}
If we know $[S]$ and $[P]$ for a substrate/phosphatase pair
(Table~\ref{tab1}), and also the enzymatic kinetic parameters
$K_M^\text{kin}$, $K_M^\text{pho}$, $\kappa_r$, $\rho_r$, we can then
find the unique value of $\gamma_K$ which makes Eq.~\eqref{t3} true.
Given a free kinase input trajectory with the corresponding fluctuation
time $\gamma_K^{-1}$, the system will behave as an optimal WK filter,
achieving the mutual information $I_\text{max} = -(1/2) \log_2 E_\wk$.
Note that Eq.~\eqref{t3} is independent of $F$, and hence it is the
the timescale of the input fluctuations (rather than their
amplitude) that is relevant for optimality.

Unfortunately we do not have precise enzymatic kinetic parameter
values for the proteins in the pathway.  As a workaround, we can
identify a physiologically realistic range from estimates available in
the literature: $K_M^\text{kin} = 10^{-8} - 10^{-6}$ M,
$K_M^\text{pho} = 10^{-8} - 10^{-6}$ M, $\kappa_r = 1 - 10$ s$^{-1}$,
$\rho_r = 0.05 - 0.5$
s$^{-1}$~\cite{Huang96,ElMasri99,Kholodenko99,Schoeberl02}.  By
drawing randomly from this range (for each parameter $p$ using a
uniform distribution of $\log_{10} p$ between the maximum and minimum
values identified above) we can find a distribution of plausible
values for $\gamma_K$.  The median of this distribution is reported in
Table~\ref{tab1} for each substrate, with the 68\% confidence
intervals in parentheses.  We also list the corresponding median and
confidence intervals for $I_\text{max}$.

The results show that $\gamma_K^{-1}$ is typically on the scale of
minutes, which means that these enzymatic loops in yeast optimally
transmit input fluctuations (driven by environmental changes) that
vary on this timescale.  The $I_\text{max}$ values fall in the range
$1.5 -2.2$ bits, which compares favorably with experimental values of
mutual information measured in other eukaryotic signaling pathways:
$0.6-1.6$ bits for mouse fibroblast cells responding to external
stimuli like tumor necrosis factor~\cite{Cheong11}.  In the
experimental case the mutual information is calculated between the
input and output of the entire pathway rather than for a single
enzymatic loop in the cascade, and hence the measured value will be a
lower bound on the information transferred through any loop in the
pathway.

\begin{figure}
\centering\includegraphics[width=\columnwidth]{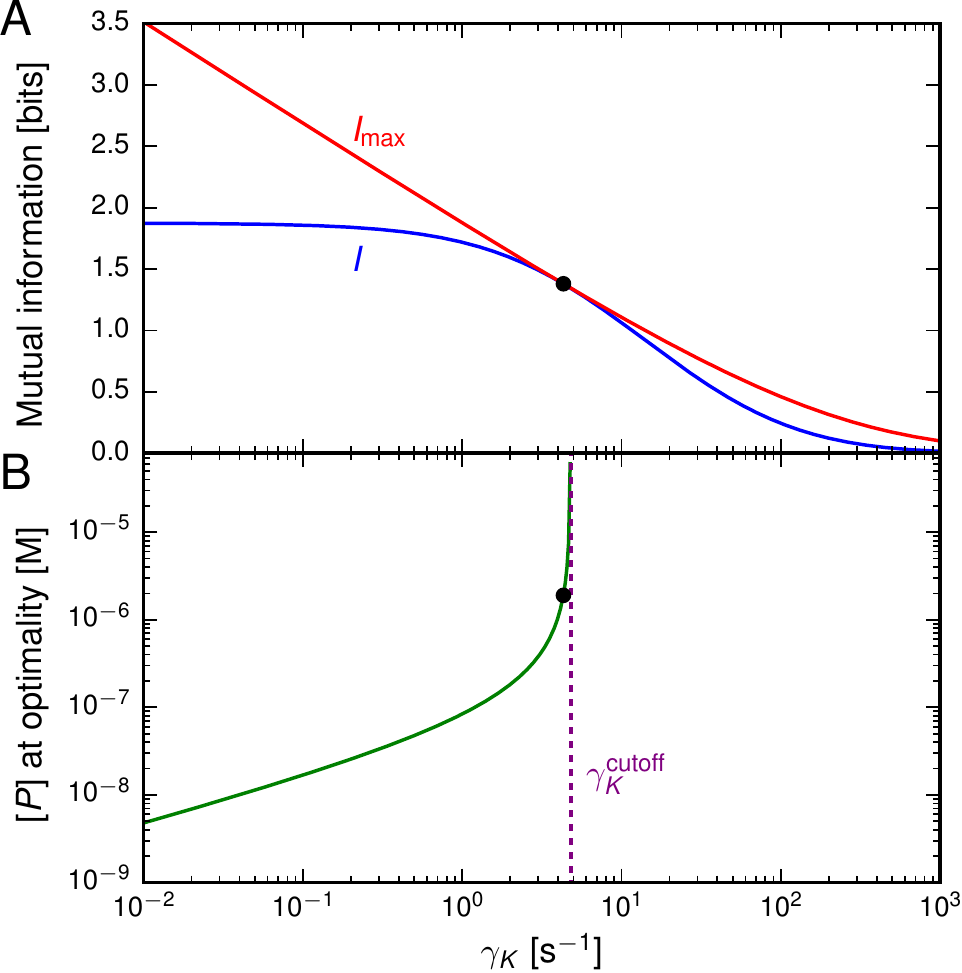}
\caption{\rev{A: $I$ (blue curve) is the mutual information for an
    enzymatic push-pull loop given various kinase input signals.  The
    input is characterized by a range of $\gamma_K$ values (and hence
    different fluctuation timescales $\gamma_K^{-1}$).  The enzymatic
    parameters are set at: $K_M^\text{kin} = K_M^\text{pho} = 0.1$
    $\mu$M, $\kappa_r=3.0$ s$^{-1}$, $\rho_r = 0.2$ s$^{-1}$.  Mean
    substrate and phosphatase concentrations are fixed at $[S] = 0.38$
    $\mu$M, $[P] = 1.9$ $\mu$M, the values for Hog1 from
    Table~\ref{tab1}.  For comparison, we also plot $I_\text{max}$
    (red curve), the mutual information if WK optimality were to be
    achieved (i.e. if Eq.~\eqref{t3} was satisfied).  When
    $\gamma_K = 4.3$ min$^{-1}$ (indicated by a dot) Eq.~\eqref{t3} is
    actually satisfied, so $I=I_\text{max}$.  B: The phosphatase
    concentration $[P]$ that is necessary for WK optimality to hold at
    each $\gamma_K$, calculated from Eq.~\eqref{t3} for the same $[S]$
    concentration and enzymatic parameters as above.  The dot marks
    $[P] = 1.9$ $\mu$M, the phosphatase concentration in panel A.  The
    dashed line is $\gamma_K^\text{cutoff}$, discussed in the text.}}\label{yeast}
\end{figure}

Now consider an enzymatic push-pull system that operates at a
particular set of $[S]$ and $[P]$ concentrations, for example Hog1
from Table~\ref{tab1}.  For concreteness, let us choose enzymatic
parameters $K_M^\text{kin} = K_M^\text{pho} = 0.1$ $\mu$M,
$\kappa_r=3.0$ s$^{-1}$, $\rho_r = 0.2$ s$^{-1}$.  With these
concentrations and parameters, Eq.~\eqref{t3} is valid only when
$\gamma_K = 4.3$ min$^{-1}$.  If an input signal has the corresponding
fluctuation timescale $\gamma_K^{-1} = 0.23$ min, the mutual
information $I = I_\text{max} = 1.4$ bits.  But what happens if this
system encounters a different input signal, with a smaller or larger
fluctuation timescale?  Intuitively one would expect that if it can
efficiently transmit fluctuations that vary on scales of $0.23$ min,
it should also work well for signals that vary on longer scales, where
$\gamma_K < 4.3$ min$^{-1}$.  Fig.~\ref{yeast}A shows what happens to
the mutual information $I$ when $\gamma_K$ is varied, but all the
other parameters and concentrations are kept fixed.  $I$ is calculated
using $I = (-1/2) \log_2 E$, with $E$ determined from Eq.~\eqref{e7}
and the mapping of Eq.~\eqref{t2}.  When $\gamma_K = 4.3$ min$^{-1}$
(marked by a dot), we have $I= I_\text{max}$.  For $\gamma_K \ne 4.3$
min$^{-1}$ we have $I < I_\text{max}$, since Eq.~\eqref{t3} no longer
holds, but the value of $I_\text{max}$ is itself dependent on
$\gamma_K$.  As expected, both $I$ and $I_\text{max}$ increase with
decreasing $\gamma_K$: it is easier to transmit signals with longer
fluctuation timescales, and despite the fact that $I$ does not achieve
optimality, it does saturate in the limit $\gamma_K \to 0$ at a larger
value than at $\gamma_K = 4.3$ min$^{-1}$.  In the opposite limit,
where $\gamma_K \to \infty$, both $I_\text{max}$ and $I$ decrease to
zero.  Thus the value of $\gamma_K$ where WK optimality is achieved
acts as an effective low-pass filter bandwidth for the system:
frequencies smaller than $4.3$ min$^{-1}$ are transmitted, while those
greater than $4.3$ min$^{-1}$ are suppressed.  This is true even if
the input signal is non-Markovian: in Ref.~\cite{Hinczewski14} we
showed a similar low-pass filtering behavior when the input signal had
a deterministic oscillatory contribution (using a sinusoidal $F(t)$
instead of a constant $F$).  This kind of deterministic input can be
implemented experimentally using
microfluidics~\cite{Mettetal08,Hersen08}, stimulating the Hog1
signaling pathway in yeast using periodic osmolyte shocks.
Ref.~\cite{Hersen08} characterized the bandwidth for the whole Hog1
pathway, including all the upstream components that amplify the signal
before it activates Hog1, and found a value of $ \approx 0.3$
min$^{-1}$.  This is consistent with our estimates for the Hog1
enzymatic loop by itself, since the bandwidth of one component in a
serial pathway will always be greater or equal to the bandwidth of the
entire pathway.

Thus the evolutionary rationale for tuning $[P]$ to satisfy
Eq.~\eqref{t3} at a particular value of $\gamma_K$ is clear: it will
allow the system to transmit signals with frequencies up to!
$\gamma_K$.  In general to get a larger bandwidth at fixed $[S]$, one
needs a larger $[P]$.  Fig.~\ref{yeast}B shows how the $[P]$ necessary
to satisfy WK optimality [Eq.~\eqref{t3}] varies with $\gamma_K$.  At
small $\gamma_K$, Eq.~\eqref{t3} scales like
$[P] \propto \gamma_K^{1/2}$: for every tenfold increase in $[P]$, the
effective bandwidth increases a hundredfold.  Thus if sufficient
bandwidth to accurately represent environmental fluctuations is
necessary for survival, there will be a strong evolutionary incentive
to increase $[P]$.  But this incentive works only up to a point.  When
the desired bandwidth $\gamma_K$ becomes so large that the denominator
of Eq.~\eqref{t3} approaches zero, at about
$\gamma^\text{cutoff}_K \approx \rho_r^2 (K_M^\text{kin} +[S])^2 /
K_M^\text{kin} \kappa_r [S]$ for $\rho_r \ll \kappa_r$, the necessary
$[P]$ to achieve optimality blows up.  For our parameters this cutoff
is $\gamma^\text{cutoff}_K \approx 4.8$ min$^{-1}$.  Given the rapidly
diminishing returns in extra bandwidth as $[P]$ approaches the
blow-up, it makes sense for evolution to stop just short of the
cutoff.  Indeed for the Hog1 case the $[S]$ and $[P]$ give a bandwidth
$\gamma_K = 4.3$ min$^{-1}$ that is $90\%$ of the cutoff value.  For
Fus3 and Slt2 that ratio is $20\%$, smaller but still within an order
of magnitude of the cutoff.  It will be interesting to check this
hypothesis across a wider set of proteins, if more detailed data
becomes available: has evolution always tuned $[P]$ relative to $[S]$
to increase bandwidth up to the point of diminishing returns?  It will
also be worthwhile in the future to consider how the above arguments
are modified in a more complex enzymatic system where activation
requires multiple phosphorylation steps.  In this case the mapping
onto the two-species system will lead to a non-zero time delay
$\alpha >0$.}

\subsubsection{Beyond the linear filter approximation} 
Ref.~\cite{Hinczewski14} also considered what happens when the
production function $R(I)$ becomes nonlinear, and we allow molecular
populations of arbitrary size (not necessarily in the continuum
limit).  For the two-species signaling pathway with $\alpha = 0$, a
rigorous analytical solution for $E$ was derived in this generalized
case.  From this solution it is clear that $E \ge E_\wk$ always
remains true, with equality only achievable when $R$ is linear.  The
mathematical forms of $E_\wk$ and $\Lambda$ remain the same in the
generalized theory, but the coefficients $\sigma_0$ and $\sigma_1$
that enter into $\Lambda$ (and hence $E_\wk$) are given by the
averages $\sigma_0 = \langle R(x(t))\rangle$ and
$\sigma_1 = \bar{x}^{-1}\langle (x-\bar{x}) R(x(t)) \rangle$.  These
reduce to the definitions of $\sigma_0$ and $\sigma_1$ given earlier
when $R$ is linear and $\alpha =0$.  Interestingly, $\sigma_1$ can be
greater than $\sigma_0 /\bar{x}$ in the nonlinear case, corresponding
to a sigmoidal production function with a steep slope near $\bar{x}$.
This can be beneficial for noise filtering, by increasing $\Lambda$
without making $\sigma_0$ larger.  Such sigmoidal production functions
have in fact been seen in certain signaling cascades, a phenomenon
known as ultrasensitivity~\cite{Goldbeter81}.  However since nonlinear
contributions to $R(I)$ always push $E$ above $E_\wk$, there is a
tradeoff between increasing $\Lambda$ through higher $\sigma_1$ and
eventually ruining the signal fidelity by making $R$ too
nonlinear~\cite{Hinczewski14}.  In any case, the bound $E_\wk$ still
applies, illustrating the usefulness of the WK approach even outside
the regimes where the continuum, linear approximation is strictly
valid.

\subsection{Three-species signaling pathway}

\subsubsection{Dynamical model}

Our second example is a three-species signaling pathway
(Fig.~\ref{net}B), extending the two-species model from the previous
section.  Since many biological signaling systems are arranged in
multiple stages~\cite{Heinrich02,Thattai02}, \rev{like the MAPK
  cascade,} progressively amplifying the signal, such a generalization
is a natural next step.  It also allows us to better understand the
origin of the time delay $\alpha$ in the two-species model, relating
it to the finite time of signal propagation when intermediate species
are involved.

We now have three molecular types, X, Y, and Z \rev{(active kinase
  populations)} with populations $x(t)$, $y(t)$, and $z(t)$ governed
by the Langevin equations:
\begin{equation}\label{e27}
\begin{split}
\frac{dx(t)}{dt} &= F - \gamma_x x(t) + n_x(t),\\
\frac{dy(t)}{dt} &= R_a(x(t)) -\gamma_y y(t) + n_y(t),\\
\frac{dz(t)}{dt} &= R_b(y(t)) -\gamma_z z(t) + n_z(t),
\end{split}
\end{equation}
where the noise functions have correlations
$\langle n_\mu(t) n_\nu(t^\prime) \rangle = 2\delta_{\mu\nu}\gamma_\mu
\bar{\mu} \delta(t-t^\prime)$ for $\mu,\nu = x,y,z$.  The linear
production functions are given by
$R_a(x) = \sigma_{a0} + \sigma_{a1} (x-\bar{x})$,
$R_b(y) = \sigma_{b0} +\sigma_{b1}(y-\bar{y})$, and the means are
$\bar{x} = F/\gamma_x$, $\bar{y} = \sigma_{a0}/\gamma_y$,
$\bar{z} = \sigma_{b0}/\gamma_z$.  Note there is no explicit time
delay in the production at each stage (the reactions involved are
assumed to be fast compared to the other timescales in the problem).
However, as we will see later, when we consider the overall signal
propagation from X to Z, the presence of Y will introduce an effective
time delay that plays the same role as $\alpha$ in the two-species
model.

\subsubsection{Mapping onto a noise filter} 
\begin{figure}
\includegraphics[width=\columnwidth]{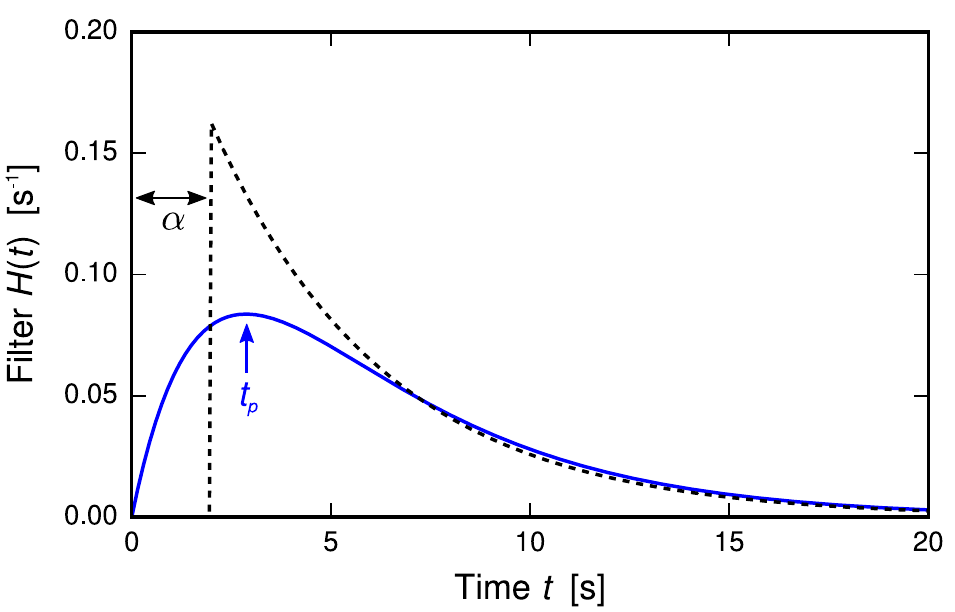}
\caption{The filter function $H(t)$ for the three-species pathway
  [Eq.~\eqref{e30}] is drawn as a solid blue curve.  The parameters are
  $\gamma_y = 0.50$ s$^{-1}$, $\gamma_z=0.23$ s$^{-1}$, $G=12.3$,
  which satisfy the WK optimality conditions of
  Eqs.~\eqref{e34}-\eqref{e35b}, with the remaining parameters set at:
  $\Lambda_a = 100$, $\Lambda_b = 2$, $\gamma_x = 0.05$ s$^{-1}$,
  $\sigma_{a1} = \sigma_{b1} = 1$ s$^{-1}$.  The peak position $t_p$
  is marked with an arrow.  For comparison, the two-species optimal
  filter $H_\wk(t)$ [Eq.~\eqref{e21}] is drawn as a black dashed curve, using the approximate mapping $\alpha = \gamma_y^{-1}$, $\Lambda = \Lambda_b \sqrt{1+\Lambda_a}$  discussed after Eq.~\eqref{e37}.}\label{filt}
\end{figure}

\rev{Since the
  three-species model is a signaling pathway, we are interested in the
  mutual information between the beginning and end of the pathway.
  The optimization problem will be analogous to Eq.~\eqref{a1}, but in
  terms of $I(\delta x; \delta z)$ rather than
  $I(\delta x; \delta y)$.  Note that the set $\Omega$ of system
  parameters over which we optimize will be different than the
  two-species model, as we will describe below. Maximizing
  $I(\delta x; \delta z)$ over $\Omega$ will be equivalent to
  minimizing the scale-independent error
  $E = \text{min}_{\tilde{A}} \epsilon(\tilde{A} \delta x(t), \delta
  z(t))$.  The gain $G$ will be the value of $\tilde{A}$ at which
  $\epsilon(\tilde{A} \delta x(t), \delta z(t))$ is minimized.} As
before, we can rewrite Eq.~\eqref{e27} in terms of deviations from
mean values, and solve the resulting system of equations in Fourier
space.  Focusing on $\delta x(\omega)$ and $\delta z(\omega)$ we find:
\begin{equation}\label{e28}
\begin{split}
\delta x(\omega) &=\frac{n_x(\omega)}{\gamma_x - i\omega},\\
\delta z(\omega) &= \frac{G^{-1}\sigma_{a1}\sigma_{b1}}{(\gamma_y - i\omega)(\gamma_z - i\omega)} \Biggl(G \delta x(\omega)\\
&\qquad+\frac{G n_y(\omega)}{\sigma_{a1}} + \frac{G(\gamma_y-i\omega) n_z(\omega)}{\sigma_{a1}\sigma_{b1}}\Biggr),
\end{split}
\end{equation}
We again have a direct mapping onto
the noise filter form of Eq.~\eqref{e6}, with:
\begin{equation}\label{e29}
\begin{split}
\tilde{s}(\omega) &= \delta z(\omega), \qquad s(\omega) = G \delta x(\omega),\\
H(\omega) &= \frac{G^{-1}\sigma_{a1}\sigma_{b1}}{(\gamma_y - i\omega)(\gamma_z - i\omega)},\\
n(\omega) &= \frac{G n_y(\omega)}{\sigma_{a1}} + \frac{G(\gamma_y-i\omega) n_z(\omega)}{\sigma_{a1}\sigma_{b1}}.
\end{split}
\end{equation}
In the time-domain the filter function $H(t)$ is:
\begin{equation}\label{e30}
H(t) = G^{-1}\sigma_{a1}\sigma_{b1}\frac{(e^{-\gamma_z t}-e^{-\gamma_yt })}{(\gamma_y-\gamma_z)} \Theta(t).
\end{equation}
For large $t$ the function $H(t)$ decays exponentially, approximately
with rate $\gamma_z$ or $\gamma_y$ depending on which is smaller.  As
$t$ decreases, $H(t)$ peaks at
$t = \log(\gamma_y/\gamma_z)/(\gamma_y-\gamma_z) \equiv t_p$ and then
goes to zero at $t=0$.  Fig.~\ref{filt} shows a representative $H(t)$
curve, and one can see that it qualitatively resembles the
time-delayed $H(t)$ of the two-species model (dashed curve) where
$t_p$ roughly plays the role of the time delay $\alpha$.  We will make
this connection between the two models concrete later.

\subsubsection{Optimality}

In addition to the characteristic timescale of the signal
$\gamma_x^{-1}$, there is now another timescale, $\gamma_y^{-1}$,
related to the deactivation of the intermediate species Y \rev{(the
  action of phosphatases on Y)}.  The Y population cannot respond to
input fluctuations on timescales much smaller than $\gamma_y^{-1}$, so
$\gamma_y^{-1}$ can also be interpreted as the characteristic response
time of Y.  This extra timescale appears in the noise function
$n(\omega)$ in Eq.~\eqref{e29}, and thus is another parameter
determining the power spectrum $P_{cc}$, along with $\gamma_x$, $F$,
and the coefficients $\sigma_{a0}$, $\sigma_{a1}$, $\sigma_{b0}$,
$\sigma_{b1}$.  \rev{These system parameters constitute the set
  $\Psi$, and determine $P_{ss}$, $P_{cs}$, and $P_{cc}$ up to an
  overall scaling factor.  The remaining set $\Omega$ again has only
  one parameter, $\gamma_z$, which is the degree of freedom that
  allows $H(t)$ to vary and possibly approach $H_\wk(t)$.  As we saw
  in the two-species case, $\gamma_z$ is effectively related to the
  kinetic parameters and populations of the phosphatase enzymes that
  dephosphorylate species Z.}

\begin{figure}
\includegraphics[width=\columnwidth]{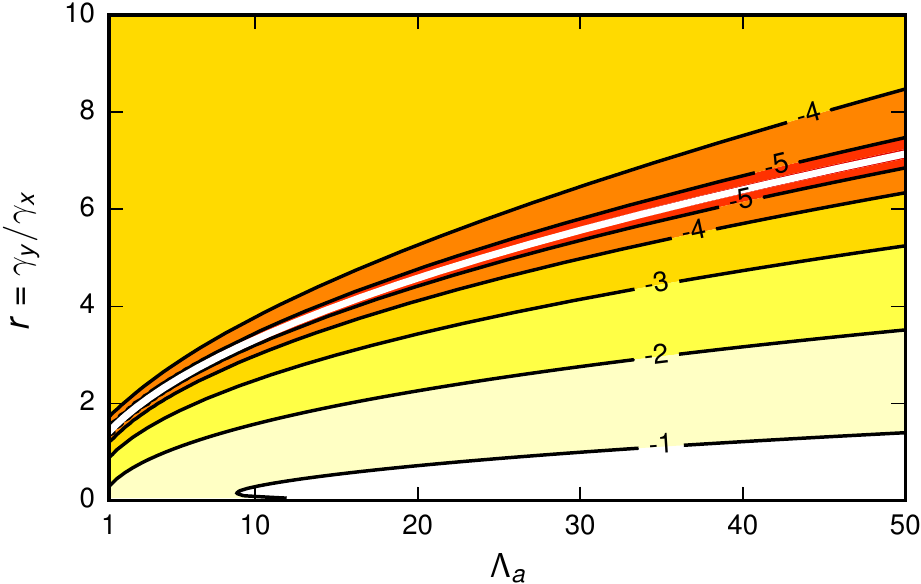}
\caption{Contour plot of $\log_{10} (E_\text{min}/E_\wk - 1)$ for the
  three-species pathway, where
  $E_\text{min} = \text{min}_{\gamma_z} E$ is the minimal achievable
  error (minimizing Eq.~\eqref{e33}) and $E_\wk$ is the WK optimum.
  The parameter $\Lambda_b = 5$ is fixed, and the minimization is
  carried out at different values of $\Lambda_a$ and
  $r = \gamma_y/\rev{\gamma_x}$.  WK optimality occurs at
  $r = \sqrt{1+\Lambda_a}$, indicated by a white
  curve.}\label{cont}
\end{figure}

The optimality calculation proceeds analogously to the two-species
case, using the $\alpha=0$ version of Eq.~\eqref{e9} since there is no
explicit time delay and we would like to optimize over the class of
all linear filters where $H(t) = 0$ for $t<0$.  The final result for the
optimal error $E_\wk$ is given by:
\begin{equation}\label{e31}
E_\wk = 1 - \frac{16 r \Lambda_a \Lambda_b}{M^2_+(r,\Lambda_a,\Lambda_b) M^2_-(r,\Lambda_a,\Lambda_b)},
\end{equation}
where $r=\gamma_y/\gamma_x$,
$\Lambda_a = \bar{x} \sigma_{a1}^2/(\gamma_x \sigma_{a0})$,
$\Lambda_b = \bar{y} \sigma_{b1}^2/(\gamma_x \sigma_{b0})$.  The
functions $M_\pm$ that appear in the denominator of Eq.~\eqref{e31}
are defined as:
\begin{equation}\label{e32}
\begin{split}
&M_\pm(r,\Lambda_a,\Lambda_b) = 2+\sqrt{2}\Bigl(1+r^2+r\Lambda_b\\
&\quad \pm \sqrt{1+r^4+2r^3 \Lambda_b -2r (1+2\Lambda_a)\Lambda_b + r^2(\Lambda_b^2-2)}\Bigr)^{1/2}.
\end{split}
\end{equation}
Unlike the two-species system, it is not always possible to tune
$\gamma_z$ such that $E = E_\wk$ exactly.  The system cannot implement
all possible $H(t)$, and as such cannot necessarily attain
$H(t) = H_\wk(t)$ by varying $\gamma_z$.  In Fig.~\ref{cont} we show a
contour plot of $\log_{10} (E_\text{min}/E_\wk - 1)$, the logarithm of
the fractional difference between the minimal achievable error,
$E_\text{min} = \text{min}_{\gamma_z} E$, and $E_\wk$.  We fix
$\Lambda_b = 5$, and perform the minimization along $\gamma_z$
numerically for a given $\Lambda_a$ and $r$.  The function $E$ we
minimize is the error for the noise filter system in
Eq.~\eqref{e29}, calculated using Eq.~\eqref{e7}:
\begin{equation}\label{e33}
E = \frac{(1+r)^2+(2+3r)q + q^2 + \frac{(1+r)^2(1+q)^2(r+q+\Lambda_b)}{\Lambda_a \Lambda_b}}{(1+r)(1+q)\left(1+r +q+ \frac{(1+r)(1+q)(r+q+\Lambda_b)}{\Lambda_a\Lambda_b}\right)},
\end{equation}
where $q = \gamma_z/\gamma_x$.  For the range of $\Lambda_a$ and $r$
shown, the largest deviation from WK optimality ($\sim 30\%$) occurs
when $r \ll 1$, corresponding to timescales $\gamma_y^{-1}$ for the Y
species that are much longer than the characteristic input signal
timescale $\gamma_x^{-1}$.  This is not a good regime for signaling,
because the response time of the intermediate species is too slow to
accurately capture the changes in the input signal.  On the other hand
for $r > 1$ the error $E_\text{min}$ is always within $14\%$ of
$E_\wk$.  In fact, for a particular curve of $r$ values, given by
\begin{equation}\label{e34}
r = \frac{\gamma_y}{\gamma_x} = \sqrt{1+ \Lambda_a}
\end{equation}
the system reaches WK optimality, with $E_\text{min} = E_\wk$.  This
curve is colored white in Fig.~\ref{cont}.  If Eq.~\eqref{e34} is
satisfied, the value of $\gamma_z$ where optimality occurs is:
\begin{equation}\label{e35}
\gamma_z = \gamma_x \sqrt{1+ \Lambda_b \sqrt{1+\Lambda_a}},
\end{equation}
and the corresponding optimal scaling factor \rev{$G_\wk$} takes the form:
\begin{equation}\label{e35b}
\rev{G_\wk} = \frac{\sigma_{a1}\sigma_{b1}(\gamma_x + \gamma_y)(\gamma_x + \gamma_z)}{\gamma_x^3\gamma_y \Lambda_a \Lambda_b}.
\end{equation}
Since Eq.~\eqref{e34} is the same as the relation between $\gamma_y$
and $\gamma_x$ in Eq.~\eqref{e22} for the two-species case, when
Eqs.~\eqref{e34}-\eqref{e35} are fulfilled the system is both an
optimal WK filter between X and Z, and also between X and Y.  If
$r > \sqrt{1+ \Lambda_a}$ (the region above the white curve in
Fig.~\ref{cont}) we can no longer exactly reach $E_\wk$, but the
difference between $E_\text{min}$ and $E_\wk$ is negligible, less than
0.1\%.  Thus the regime of large $r$ (fast response timescales
$\gamma_y^{-1}$ for species Y) is generally favorable for noise
filtering.  This intuitively agrees with our expectation: if the
dynamics of the intermediate species is fast, it will not impede
signal propagation.

Let us focus on the parameter subspace where the system is WK optimal,
and hence Eqs.~\eqref{e34}-\eqref{e35} hold for $\gamma_y$ and
$\gamma_z$.  Using these conditions $E_\wk$ from Eq.~\eqref{e31} can
be rewritten in a simpler form:
\begin{equation}\label{e36}
E_\wk = 1 - \frac{\Lambda_a \Lambda_b \sqrt{1+\Lambda_a}}{(1+\sqrt{1+\Lambda_a})^2 (1+ \sqrt{1+\Lambda_b\sqrt{1+\Lambda_a}})^2}.
\end{equation}
Consider this equation in two limiting cases: 

\begin{enumerate}

\item Assume that $\gamma_y^{-1} \ll \gamma_z^{-1}$, so the Z response
  time is much slower than Y.  From Eqs.~\eqref{e34}-\eqref{e35} this
  is equivalent to assuming that
  $\Lambda_a \gg \Lambda_b\sqrt{1+\Lambda_a}$.  In this limit
  Eq.~\eqref{e36} can be expanded to lowest order in $\Lambda_a^{-1}$:
\begin{equation}\label{e37}
E_\wk = \frac{2}{1+\sqrt{1+\Lambda}} + \frac{2\Lambda \Lambda_a^{-1/2}}{(1+\sqrt{1+\Lambda})^2} +{\cal O}(\Lambda_a^{-1}),
\end{equation}
where we have introduced an effective parameter
$\Lambda = \Lambda_b \sqrt{1+\Lambda_a}$, which makes clear that the
result has the same structure as Eq.~\eqref{e25} for the two-species
$E_\wk$ with small $\alpha$.  The role of $\alpha\gamma_x$ is played
by $\Lambda_a^{-1/2}$.  For large $\Lambda_a$ this can be approximated
as $\Lambda_a^{-1/2} \approx \gamma_x/\gamma_y$ using Eq.~\eqref{e34}.
Hence the timescale $\gamma_y^{-1}$ acts like an effective time delay
$\alpha$.  When the Y response time is fast, the three-species pathway
can be mapped onto a two-species model with a small time delay
$\alpha = \gamma_y^{-1}$ and $\Lambda = \Lambda_b \sqrt{1+\Lambda_a}$.
The two species in this reduced model are X and Z, with the optimality
condition $\gamma_z = \gamma_x \sqrt{1+\Lambda}$.  The influence of
the Y is encoded in the time delay, and also in the factor
$\sqrt{1+\Lambda_a}$ renormalizing $\Lambda_b$ in the expression for
$\Lambda$.

\item Assume that $\gamma_y^{-1} \gg \gamma_z^{-1}$, so the Y response
  time is much slower than Z.  An analogous calculation shows that
  Eq.~\eqref{e36} can be written in the form of Eq.~\eqref{e25}, with
  an effective $\Lambda = \Lambda_a$ and $\alpha = \gamma_z^{-1}$.
\end{enumerate}

Both these results are also consistent with the argument above
identifying the peak position $t_p$ of the filter $H(t)$ with the
rough value of $\alpha$.  When $\gamma_y^{-1} \ll \gamma_z^{-1}$, we
find
$t_p = \log(\gamma_y/\gamma_z)/(\gamma_y-\gamma_z) \sim \gamma_y^{-1}
= \alpha$ up to logarithmic corrections.  Similarly when
$\gamma_y^{-1} \gg \gamma_z^{-1}$, we find
$t_p \sim \gamma_z^{-1} = \alpha$ up to logarithmic corrections.

\section{Realizing a noise filter in a negative feedback loop}

\rev{We will now consider WK filter theory in a different biological
  context, with a different formulation of the optimization problem.}
The system is a simple two-species negative feedback loop
(Fig.~\ref{net}C), where X catalyzes the activation of Y, and Y
promotes the deactivation of X.  Such negative feedback is a
widespread phenomenon in biological reaction
networks~\cite{Becskei00,Thattai01,Simpson03,Austin06,Dublanche06,Cox06,Zhang09,Singh09},
and is capable of suppressing noise in the sense of reducing the
fluctuations of the inhibited species X.  This kind of noise
suppression is conceptually different than the noise filtering during
signal propagation we described in the previous two examples.  But as
we will see shortly, once we construct a mapping onto the noise filter
formalism, the mathematical structure of the optimization results is
remarkably similar to those of the signaling pathways.

\subsubsection{Dynamical model}

The Langevin equations for the populations $x(t)$ and $y(t)$ are:
\begin{equation}\label{e38}
\begin{split}
\frac{dx(t)}{dt} &= \Phi(y(t))-  \gamma_x x(t) + n_x(t),\\
\frac{dy(t)}{dt} &= R(x(t)) -\gamma_y y(t) + n_y(t).\\
\end{split}
\end{equation}
The production function $R$ is linearized as before:
$R(x) = \sigma_{0} + \sigma_{1} (x-\bar{x})$.  The X production
function $\Phi$ is dependent on Y, and for small fluctuations $y(t)$
near $\bar{y}$ can be linearized as: $\Phi(y) = F - \phi(y-\bar{y})$,
where $\phi \ge 0$ represents the strength of the negative feedback on
X.  The means are $\bar{x} = F/\gamma_x$,
$\bar{y} = \sigma_0/\gamma_y$.  We have no explicit time delay in the
production or feedback (though it can be added to the theory in a
straightforward way).  As with the three-species pathway, an effective
time delay will arise naturally out of the dynamics, related to the
characteristic response time $\gamma_y^{-1}$ of the species Y that
mediates the feedback.

\subsubsection{Mapping onto a noise filter}

The solutions for the deviations $\delta x(\omega)$ and $\delta y(\omega)$ in Fourier space are:
\begin{equation}\label{e39}
\begin{split}
\delta x(\omega) &= \frac{(\gamma_y-i\omega)n_x(\omega) - \phi n_y(\omega)}{\phi \sigma_1+(\gamma_x - i\omega)(\gamma_y-i\omega)},\\
\delta y(\omega) &= \frac{(\gamma_x-i\omega)n_y(\omega) + \sigma_1 n_x}{
\phi \sigma_1+(\gamma_x - i\omega)(\gamma_y-i\omega)}.
\end{split}
\end{equation}
To map this to a noise filter we will follow the approach of
Ref.~\cite{Hinczewski16}: 
\begin{equation}\label{e40}
s(\omega) = \delta x(\omega)|_{\phi=0}, \qquad \tilde{s}(\omega) = \delta x(\omega)|_{\phi=0} - \delta x(\omega).
\end{equation}
The signal is thus $\delta x(\omega)$ in the absence of feedback
($\phi=0$), and the estimate is the difference between this result and
the $\delta x(\omega)$ in the presence of feedback.  With these
definitions, $\delta x(\omega)$ can be decomposed into two parts:
\begin{equation}\label{e41}
\delta x(\omega) = s(\omega) - \tilde{s}(\omega) = s(\omega) - H(\omega)(s(\omega) + n(\omega)),
\end{equation}
where comparison with Eq.~\eqref{e39} allows us to identify:
\begin{equation}\label{e42}
H(\omega) = \frac{\phi \sigma_1}{\phi \sigma_1 + (\gamma_x - i \omega)(\gamma_y - i\omega)}, \quad n(\omega) = \frac{n_y(\omega)}{\sigma_1}.
\end{equation}
The motivation behind this mapping is that negative feedback damps the
fluctuations $\delta x$, which in the filter formalism is equivalent
to making the estimate $\tilde s$ similar to $s$.  The relative mean
squared error from Eq.~\eqref{e1} has a simple interpretation in this
case,
\begin{equation}\label{e43}
\epsilon(s(t),\tilde{s}(t)) = \frac{\langle (\tilde{s}(t) - s(t))^2 \rangle}{\langle s^2(t) \rangle} = \frac{\langle (\delta x(t))^2 \rangle}{\langle (\delta x(t)|_{\phi=0})^2 \rangle},
\end{equation}
equal to the ratio of the mean squared X fluctuations with and without
feedback.  The goal of optimization is to tune $H(\omega)$ such the
$\epsilon$ is minimized, finding the feedback form that gives the
largest fractional reduction in X fluctuations.  \rev{As we discussed
  in Sec.~\ref{nfop}, minimizing $\epsilon(s(t),\tilde{s}(t))$ is
  equivalent to minimizing the scale-independent error $E$ from
  Eq.~\eqref{e3}, with $E = \epsilon$ at optimality.  This in turn is
  equivalent to maximizing the mutual information $I(s,\tilde{s})$.
  Note that the $I$ here has a more abstract interpretation than for
  signaling pathway cases, where $s(t)$ was an input into the pathway
  and $\tilde{s}(t)$ was the output from the pathway.  Here $s(t)$ is
  a hypothetical trajectory (the fluctuations $\delta x(t)|_{\phi=0}$
  which the system would exhibit if feedback was turned off), and
  $\tilde{s}(t) = \delta x(t)|_{\phi=0} - \delta x(t)$ is the
  difference between that hypothetical trajectory and the actual
  trajectory $\delta x(t)$ with feedback.  Thus $\tilde{s}$ represents
  the net effect of adding feedback into the system.  High similarity
  between $\tilde{s}$ and $s$, which translates into high mutual
  information $I(s,\tilde{s})$, corresponds to an actual trajectory
  $\delta x(t)$ that remains close to zero.  The larger the mutual
  information between $s$ and $\tilde{s}$, the more effectively the
  negative feedback can cancel $s$, keeping the fluctuations of X as
  small as possible. Though the noise filter mapping for the
negative feedback system is qualitatively different from the one
we used in the pathway examples, we will see that the results are
closely related in mathematical structure.}

We can rewrite $H(\omega)$ from Eq.~\eqref{e42} as:
\begin{equation}\label{e44}
H(\omega) = -\frac{\phi \sigma_1}{(\omega + i \lambda_+)(\omega + i \lambda_-)},
\end{equation}
where $-i \lambda_\pm$ are the roots of the denominator of
$H(\omega)$, with
$\lambda_\pm = (\gamma_x + \gamma_y \pm \sqrt{(\gamma_y-\gamma_x)^2 -
  4 \phi \sigma_1})/2$.  When the Y response time is fast,
$\gamma_y^{-1} \ll \gamma_x^{-1}$, the regime where the negative
feedback is most efficient (as we will see below), we can approximate
$\lambda_\pm$ as:
\begin{equation}\label{e45}
\lambda_+ = \gamma_y - \frac{\phi \sigma_1}{\gamma_y} + {\cal O}(\gamma_y^{-2}), \quad \lambda_- = \gamma_x + \frac{\phi \sigma_1}{\gamma_y} + {\cal O}(\gamma_y^{-2}).
\end{equation}
Note that $\lambda_\pm > 0$ in this regime, with
$\lambda_+ \gg \lambda_-$.  The corresponding time-domain filter
function $H(t)$ is:
\begin{equation}\label{e46}
H(t) = \phi\sigma_1 \frac{(e^{-\lambda_- t}-e^{-\lambda_+ t })}{(\lambda_+ - \lambda_-)} \Theta(t),
\end{equation}
which has the same structure as $H(t)$ for the three-species pathway
in Eq.~\eqref{e30}.  Just as in that case (Fig.~\ref{filt}), it is dominated by exponential decay at large $t$ (with rate constant
$\lambda_-$), then peaks at
$t_p = \log(\lambda_+/\lambda_-)/(\lambda_+-\lambda_-)$ and goes to
zero at small $t$.  The region where $H(t)$ is small, $t\le t_p$,
roughly corresponds to having a time delayed filter with
$\alpha \sim t_p$.  When $\lambda_+ \gg \lambda_-$ we have
$t_p \sim \lambda_+^{-1} \sim \gamma_y^{-1}$ up to logarithmic
corrections, so the effective time delay is controlled by the response
time of the Y species.  This makes sense because it is the Y species
that mediates the negative feedback.  More complicated models with
additional species in the feedback loop would include additional
contributions to the effective delay.

\subsubsection{Optimality} 

The optimality calculation will use the $\alpha=0$ version of
Eq.~\eqref{e9}, since there is no explicit time delay in the model.
This means we are optimizing over the class of all linear filters
where $H(t) = 0$ for $t<0$.  The relevant spectra are given by:
\begin{align}\label{e47}
P_{ss}(\omega) &= \frac{2 F} {\gamma_x^2+ \omega^2},& P_{nn}(\omega) &= \frac{2 \sigma_0}{\sigma_1^2},\nonumber\\
P_{cs}(\omega) &= P_{ss}(\omega), & P_{cc}(\omega) &= P_{ss}(\omega) + P_{nn}(\omega).
\end{align}
\rev{The spectra depend on $\Psi = \{F$, $\gamma_x$, $\sigma_0$, $\sigma_1\}$, so
the remaining degrees of freedom to optimize $H(t)$ are $\Omega = \{\phi$,
$\gamma_y\}$.}  Eq.~\eqref{e47} has the same form as Eq.~\eqref{e16}
for the two-species case (with no scaling factor $G$), and hence the
optimal filter result is the same as Eq.~\eqref{e21} with $\alpha =0$:
\begin{equation}\label{e48}
  H_\wk(t) = (\sqrt{1+\Lambda}-1) \gamma_x e^{-\gamma_x t \sqrt{1+\Lambda}}\Theta(t),
\end{equation}
where $\Lambda = \bar{x}\sigma_1^2/(\gamma_x \sigma_0)$.  The corresponding optimal error $E_\wk$ is:
\begin{equation}\label{e49}
E_\wk = \frac{2}{1+\sqrt{1+\Lambda}}.
\end{equation}
Comparing Eqs.~\eqref{e48} and \eqref{e46}, using the root expressions
in Eq.~\eqref{e45}, it is clear that $H(t) \to H_\wk(t)$ when:
\begin{equation}\label{e50}
  \gamma_y \to \infty, \qquad \phi = \frac{\gamma_y\gamma_x}{\sigma_1} (\sqrt{1+\Lambda}-1) \to \infty.
\end{equation}
Thus optimal noise filtering requires a vanishing Y response time
$\gamma_y^{-1} \to 0$, and a correspondingly large feedback strength
$\phi \propto \gamma_y$ with the proportionality constant in
Eq.~\eqref{e50}.  For any actual system $\gamma_y^{-1}$ cannot be
exactly zero, so we can check how close $E$ can get to $E_\wk$ when
the Y response time is finite.  Using Eq.~\eqref{e7}, the $E$ for the
negative feedback system is:
\begin{equation}\label{e51}
\begin{split}
E &= 1+\frac{\gamma_x(\gamma_x+\gamma_y)}{2\gamma_x(\gamma_x+\gamma_y)+\phi \sigma_1}\\
&\qquad  -\frac{\gamma_x(\gamma_x+\gamma_y)(1+\Lambda)}{2\gamma_x^2(1+\Lambda)+\gamma_x \gamma_y(2+\Lambda)+\phi\sigma_1}.
\end{split} 
\end{equation}
The error reaches its minimum $E_\text{min}$ as a function of $\phi$
when:
\begin{equation}\label{e52}
\phi = \frac{\gamma_x \gamma_y}{\sigma_1}\left[\left(\frac{2\gamma_x}{\gamma_y}+1\right)\sqrt{1+\Lambda}-1\right],
\end{equation}
with the minimum value given by:
\begin{equation}\label{e53}
\begin{split}
E_\text{min} &= \frac{2\left(1+\frac{2\gamma_x}{\gamma_y}\right)^{-1}}{1+\sqrt{1+\Lambda}} + \frac{(\Lambda-8)\left(2+\frac{\gamma_y}{\gamma_x}\right)^{-1}}{\Lambda-2(1+\sqrt{1+\Lambda})}.
\end{split}
\end{equation}
To lowest order in $\gamma_y^{-1}$, the difference between $E_\text{min}$ and $E_\wk$ is:
\begin{equation}\label{e54}
E_\text{min}-E_\wk = \frac{\Lambda \gamma_y^{-1} \gamma_x}{(1+\sqrt{1+\Lambda})^2} + {\cal O}(\gamma_y^{-2}).
\end{equation}
Eq.~\eqref{e54} has the same form as the $\alpha$ correction in
Eq.~\eqref{e25} for the two-species $E_\wk$, with an effective time
delay $\alpha = \gamma_y^{-1}/2$.  This is consistent with our
analysis above of $H(t)$, which resembles a predictive filter with
$\alpha \sim \gamma_y^{-1}$.  Since the optimal $H_\wk$ is a noise
filter with $\alpha = 0$, the difference between $H(t)$ and $H_\wk$
only vanishes when $\gamma_y^{-1} \to 0$.  For any finite
$\gamma_y^{-1}$, the performance of the optimal predictive filter
($\alpha >0$) is always worse than the optimal causal filter
($\alpha =0$) that integrates the time series up until the current
moment.

The optimal error $E_\wk$ in Eq.~\eqref{e49} satisfies the rigorous
lower-bound calculated by Lestas, Vinnicombe, and
Paulsson~\cite{Lestas10},
\begin{equation}\label{e55}
E_\wk \ge \frac{2}{1+\sqrt{1+4\Lambda}},
\end{equation}
which assumes a linear production function $R$, but allows arbitrary
negative feedback, including both linear and nonlinear cases.  In the
limit of large populations where the Langevin approach works and the
fluctuations are Gaussian, the linear filter is in fact optimal among
all filtering mechanisms, and $E_\wk$ should be the true bound on the
error.  Ref.~\cite{Hinczewski16} investigated the validity of this bound
outside the continuum approximation, and found that for a master
equation model of a negative feedback loop (based on an experimental
yeast gene circuit~\cite{Nevozhay09})
the WK bound still holds.  Moreover, the theory predicted that the
experimental circuit could be tuned to approach WK optimality (up to
corrections due to finite response times) by changing the
concentration of an extracellular inducer molecule.

\section{Conclusion}

In the three systems we have considered, WK theory provided a unified
framework for exploring the nature of information transfer in
biochemical reaction networks.  It allowed us to decompose the network
behavior into signal, noise, and filter components, and then see how
the filter properties could be tuned through the system parameters.
Since biological information processing is never instantaneous, but
always depends on the number, type, and response times of the molecular
species involved in the transmission, the filters in realistic cases
are predictive: they effectively attempt to estimate the current
true signal using a time-delayed, noisy signal trajectory from the
past.  Integrating over the past trajectory suppresses noise (at a
high biochemical cost in terms of producing signaling molecules), but
the errors due to time delay can never be completely removed.

\rev{The noise filters described here cover just a small subset of the
  diverse strategies involved in cellular decision-making, the
  mechanisms by which cells process and respond to environmental
  information~\cite{Kobayashi10,Bowsher14}.  Recently Becker, Mugler,
  and ten Wolde~\cite{Becker15} applied WK theory to {\it E.~coli}
  chemotaxis signaling networks, focusing on the capacity of cells to
  predict future environmental changes based on the current, noisy
  data transmitted through membrane receptors activated by bound
  ligands.  For Markovian extracellular signals (exponential
  autocorrelations), the theory is mathematically analogous to the
  delayed two-species signaling pathway discussed above.  They also
  showed how optimality could be achieved even when the input signal
  has long-range, non-Markovian correlations, though the optimal
  filter implementation requires a multi-layer reaction network.  More
  broadly, one can also consider noise filtering for non-stationary
  signals, where the optimal linear solution can be recursively
  derived using the powerful Kalman-Bucy filter
  approach~\cite{Kalman60,Kalman61}.  Andrews, Yi, and Iglesias showed
  how {\it E.~coli} chemotaxis can be modeled through this approach,
  and Zechner {\it et al.}~\cite{Zechner16} implemented the
  Kalman-Bucy filter in two synthetic biochemical circuits: an {\it in
    vitro} DNA strand displacement system, and an optogenetic circuit
  engineered into a living {\it E.~coli} bacterium.}  These are just
the latest iterations of a fascinating story that began with an
impractical anti-aircraft device that never shot down a single plane.

\section*{Acknowledgments}

The authors would like to thank Dave Thirumalai, Shishir Adhikari,
Tenglong Wang, Benjamin Kuznets-Speck, and Joseph Broderick for useful
discussions.


\end{document}